\documentclass[11pt]{article}
\usepackage{graphicx}
\usepackage{hyperref}

\newcommand{\qbar}{{\bar{q}}}
\newcommand{\Qbar}{{\,\overline{\!Q}{}}}
\newcommand{\Bbar}{{\,\overline{\!B}{}}}
\newcommand{\Nbar}{{\,\overline{\!N}{}}}
\newcommand{\bbar}{{\bar{b}}}
\newcommand{\bea}{\begin{eqnarray}}
\newcommand{\eea}{\end{eqnarray}}
\newcommand{\beq}{\begin{equation}}
\newcommand{\eeq}{\end{equation}}
\newcommand{\grad}{\nabla}

\newcommand{\tnc}{{\widetilde{N}_c}}
\newcommand{\vevm}{\langle M\rangle}
\newcommand{\del}{\partial}

\newcommand{\tm}{{\widetilde{m}}}
\newcommand{\Tr}{\mbox{Tr}}
\newcommand{\tr}{\mbox{tr}}
\newcommand{\Bh}{{\,\widehat{\!B}{}}}
\newcommand{\Bbarh}{{\,\widehat{\overline{\!B}}{}}}
\newcommand{\Mh}{{\,\widehat{\!M}{}}}

\begin{document}

\begin{titlepage}
\begin{center}
\today     \hfill    LBNL-53793 \\
~{} \hfill UCB-PTH-03/23  \\
~{} \hfill hep-ph/0309224\\

\vskip .2in

{\Large \bf Supersymmetric Color Superconductivity}\footnote{This work
  was supported in part by the U.S.  Department of Energy under
  Contracts DE-AC03-76SF00098, in part by the National Science
  Foundation under grant PHY-0098840. The work of HM was also
  supported by Institute for Advanced Study, funds for Natural
  Sciences.}

\vskip 0.3in

Roni Harnik and Daniel T. Larson

\vskip 0.05in
{\em Theoretical Physics Group\\
     Ernest Orlando Lawrence Berkeley National Laboratory\\
     University of California, Berkeley, 
     CA 94720, USA}

\vskip 0.05in

{\em Department of Physics, University of California\\
     Berkeley, CA 94720, USA}

\vskip 0.2in

Hitoshi Murayama\footnote{On leave of absence from Department
of Physics, University of California, Berkeley, CA 94720,
USA.}

\vskip 0.05in
{\em School of Natural Sciences, Institute for Advanced Study\\
Einstein Dr, Princeton, NJ 08540, USA}
\end{center}

\vskip .05in

\begin{abstract}
Recent interest in novel phases in high density QCD motivates the
study of high density supersymmetric QCD (SQCD), where powerful exact
results for supersymmetric gauge theories can be brought to bear in
the strongly coupled regime.  We begin by describing how a chemical
potential can be incorporated into a supersymmetric theory as a
spurion vector superfield.  We then study supersymmetric $SU(N_c)$
gauge theories with $N_f$ flavors of quarks in the presence of a
baryon chemical potential $\mu$, and describe the global symmetry
breaking patterns at low energy. Our analysis requires $\mu < \Lambda$
and is thus complementary to the variational approach that has been
successful for $\mu\gg\Lambda$. We find that for $N_f < N_c$ a
modified $U(1)_B$ symmetry is preserved, analogous to the
non-supersymmetric 2SC phase, whereas for $N_f=N_c$ there is a
critical chemical potential above which the $U(1)_B$ is broken, as it
is in the non-supersymmetric CFL phase. We further analyze the cases
with $N_c+1\le N_f <\frac32 N_c$ and find that baryon number is broken
dynamically for $\mu>\mu_c$. We also give a qualitative description of
the phases in the `conformal window', $\frac32 N_c<N_f<3N_c$, at
finite density.
\end{abstract}

\end{titlepage}

\newpage

\section{Introduction}

In the past several years there has been a lot of interest in QCD at
high density where novel phases of matter are predicted to appear (see
the reviews~\cite{Rajagopal:2000wf, Alford:2001dt} and references
therein). High densities imply that the Fermi surface lies at a high
energy scale for QCD, which means it is weakly coupled.  The tools of
BCS theory can then be used to demonstrate the instability of the
Fermi surface and the existence of lower energy condensates which
exhibit interesting symmetry breaking patterns such as color
superconductivity and color-flavor locking. These new phases might
eventually be observable in compact star dynamics.  One important
consequence of high density QCD is the dynamical breakdown of baryon
number for three or more quark flavors.

Because of asymptotic freedom previous treatments of finite density
QCD were limited to very high density where the physics is
perturbative. It would be interesting to find QCD-like theories that
are calculable even at low densities where the Fermi surface lies
within the strongly coupled regime. During the last decade, there has
been significant progress uncovering exact results in supersymmetric
gauge theories.  In particular, the low energy description of
supersymmetric-QCD (SQCD) has been described for all numbers of colors
$N_c$ and fundamental matter flavors, $N_f$. (For reviews,
see~\cite{Intriligator:1995au,Peskin:1997qi}.)

In this paper we use the exact results provided by supersymmetry to
study the effects of a baryon chemical potential on the vacuum of
supersymmetric QCD. The tight restrictions imposed by supersymmetry
allow us to reduce the set of possible vacua and symmetry breaking
patterns.  A clear limitation of this approach is that the symmetry
breaking patterns of supersymmetric and non-supersymmetric QCD are
different even in the absence of a chemical potential, so we cannot
expect them to be the same in the presence of a chemical potential. On
the other hand, one can ask a less specific question: is baryon number
broken dynamically?

The main result of our study is that indeed this question has the same
answer in both non-supersymmetric and supersymmetric theories for $N_f
\leq \frac{3}{2} N_c$ as described below, giving additional 
support to the variational results using the BCS states.  We emphasize
that the dynamical breakdown of $U(1)_B$ we have studied differs from
the simple Bose--Einstein condensation which necessarily occurs in a
theory containing scalars when the chemical potential exceeds the mass
of the scalar.  We show explicitly that baryon number is broken as a
consequence of the strong gauge dynamics at a lower chemical
potential.  It is also intriguing that baryon number can be
dynamically broken even if the chemical potential is much smaller than
the dynamical scale.

The results for the symmetry breaking patterns for non-supersymmetric
QCD in the presence of a large chemical potential have been studied in
\cite{Schafer:1999fe,Schafer:2000tw} for $N_c=3$ and are briefly 
summarized in Table~\ref{tab:qcd}. Note that for $N_f<N_c$ a
global $U(1)$ remains unbroken. For $N_f=2$ this $U(1)$ is a
combination of the original baryon number and a diagonal color
generator. However, for $N_f\ge N_c$ baryon number is spontaneously
broken. Below we will compare these
results with the possible symmetry breaking patterns in SQCD for
$N_c\ge 3$ and various numbers of flavors.
\newpage
\begin{table}[tp]
\begin{center}
\begin{tabular}{|c|l|}
\hline
$N_f$ & Unbroken Global Symmetry \\ \hline
1 & $U(1)_B$ \\
2 & $SU(2)_L\times SU(2)_R \times U(1)_B$\ \ \ \ \ \ \ \ (2SC) \\
3 & $SU(3)_{L+R+c}$ \ \ \ \ \ \ \ \ \ \ \ \ \ \ \ \ \ \ \ \ \ \ \ \ \
\ (CFL)\\
4 & $SU(2)_V \times SU(2)_V \times SU(2)_A$ \\
5 & $SU(2)_{L+R}$ \\
6 & $SU(3)_{L+R+c} \times SU(2)_{L+R+c} \times U(1)_V \times U(1)_A$ \\
\hline
\end{tabular}
\caption{Unbroken subgroup of the original $SU(N_f)_L\times SU(N_f)_R
\times U(1)_B$ global symmetry for QCD with $N_c=3$ according to
\cite{Schafer:1999fe,Schafer:2000tw}.\setcounter{footnote}{2}
\protect\footnotemark}
\label{tab:qcd}
\end{center}
\end{table}
\footnotetext{The result for $N_f=4$ in
the published version of~\cite{Schafer:1999fe} is incorrect, but has
been corrected in the e-print version. For $N_f=6$ the remaining
$U(1)_V\times U(1)_A$ symmetry is a subgroup of $SU(6)_L\times
SU(6)_R$ and is thus physically distinct from $U(1)_B$.} 

We want to study the dynamics of supersymmetric QCD in the presence of
a chemical potential $\mu$ which explicitly breaks supersymmetry.  One
main difference between the supersymmetric and non-supersymmetric
gauge theories is of course the presence of scalar quarks.  If the
chemical potential is larger than the scalar mass the squarks
immediately undergo the standard Bose--Einstein condensation. Since we
are interested in the effect of strong gauge dynamics, we will add a
stabilizing mass to prevent such condensation. Such a mass can be
either supersymmetric or SUSY-breaking. We will study both cases in
turn.

Throughout we will assume that $\mu < \Lambda$ so that the chemical
potential can be treated as a small perturbation compared to the
strong supersymmetric gauge dynamics that have been well studied.
This is a very different regime from the range of validity of the BCS
variational method in color superconductivity, namely $\mu \gg
\Lambda$.  In the latter case, the dynamics is weakly coupled and the
analysis is under control.  On the other hand, in our supersymmetric
analysis the dynamics is strongly coupled.  The added constraints from
supersymmetry allow us to draw interesting conclusions about symmetry
breaking patterns. Our analysis is therefore complementary to the
variational method of non-supersymmetric color
superconductivity. However, we cannot say anything about gauge
non-invariant quantities such as $\langle qq \rangle$. Furthermore, we
have assumed that the vacuum is translationally invariant, which means
we are ignoring the possibility of a crytalline
phase~\cite{Alford:2000ze}.
 
\begin{figure}[tbp]
  \centering
  \includegraphics[width=0.45\textwidth]{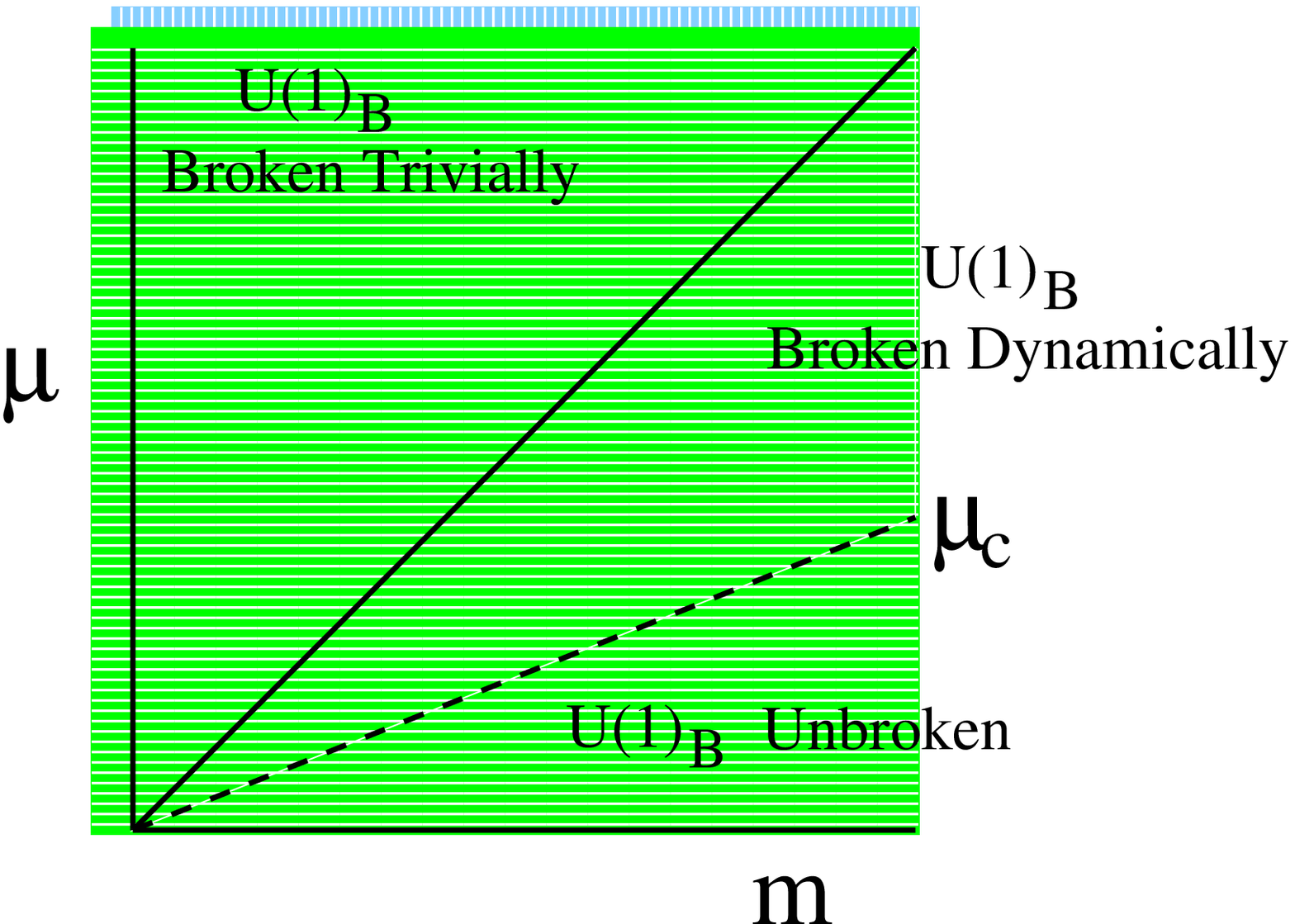}
  \includegraphics[width=0.45\textwidth]{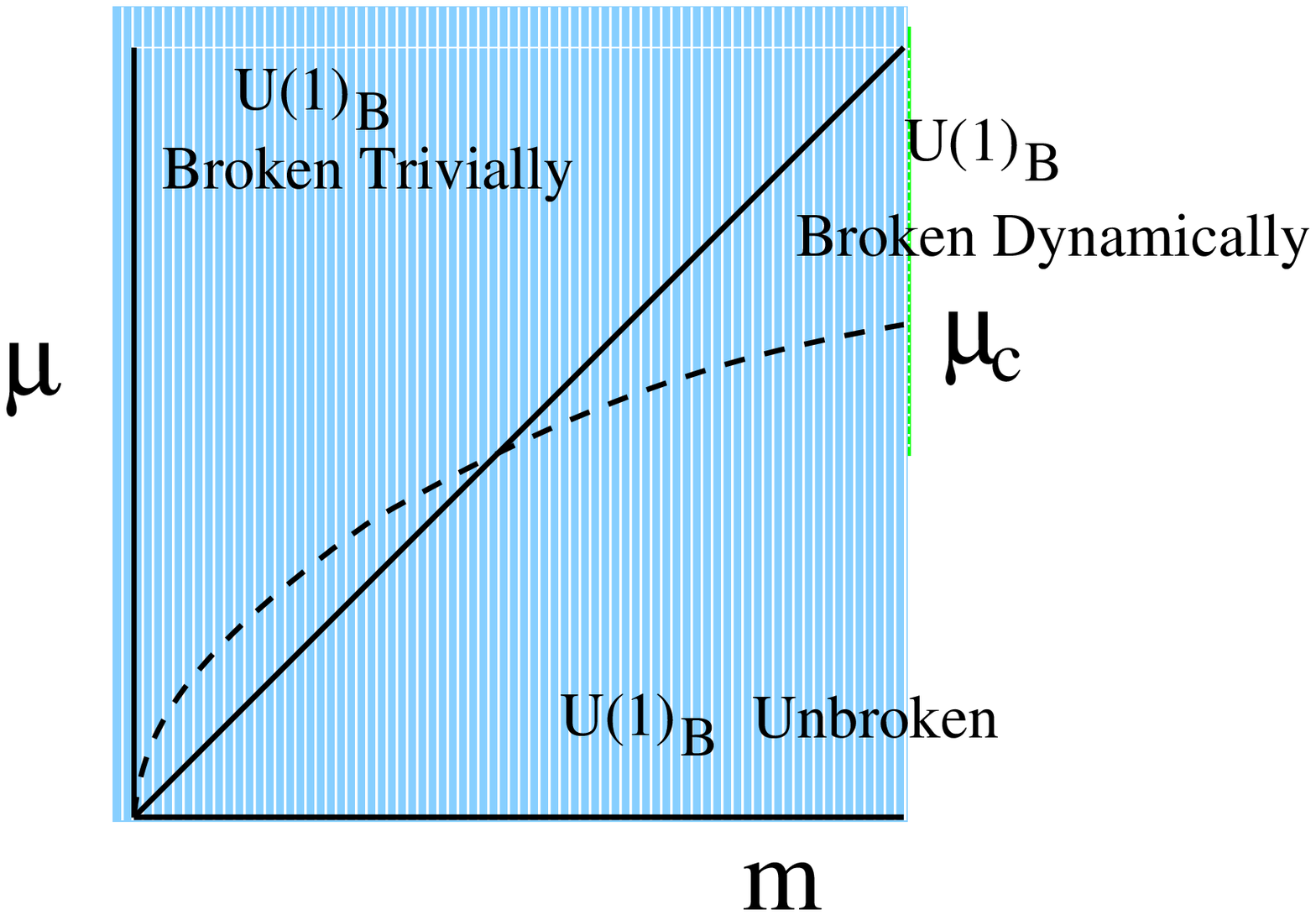}
  \caption{Schematic phase diagrams in the $\mu$--$m$ plane where $m$
  is the supersymmetric mass. The diagram on the left shows the
  situation for $N_f=N_c$ and the diagram on the right shows the
  situation for $N_c+1 \le N_f < 2N_c$. The dashed line is the
  critical chemical potential. These diagrams are only valid for $\mu
  < \Lambda$.}
\label{fig:mum-plot}
\end{figure}

Here we briefly summarize our results. We find that baryon number
is unbroken when $N_f < N_c$, but that there is a critical chemical
potential above which baryon number is broken when $N_c \le N_f <
\frac32 N_c$. This pattern matches the non-supersymmetric results. The
dependence of the critical chemical potential on the stabilizing mass
can be of two qualitatively different types, as illustrated in
Figure~\ref{fig:mum-plot}. The situation for $N_f=N_c$ is shown in the
left diagram, where $\mu_c$ is always less than the stabilizing mass
$m$. The situation for more flavors is illustrated in the right
diagram, where we see that there is a minimum value of the stabilizing
mass that allows for a nontrivial critical chemical potential. In all
cases we require $\mu < \Lambda$ so that our supersymmetric analysis
is valid.

The outline of the paper is as follows. In the next section we review
the formalism for constructing an effective Lagrangian in the presence
of a chemical potential. In Section~3 we adapt this formalism to
SQCD. In Sections~4-7 we determine the global symmetries of SQCD with
various numbers of flavors, and in Section~8 we conclude. Appendix~A
contains a simple and explicit example in quantum mechanics, as a
check on the method of including a chemical potential, while
Appendix~B reviews the exact results for soft masses in SUSY theories.

\section{Relativistic Bose-Einstein Condensation} \label{relBE}

Let us begin by reviewing the relativistic formulation of
Bose-Einstein condensation for a non-supersymmetric scalar field
theory. A nice description is given in~\cite{Kapusta, Haber:1981ts}.
There are two purposes to this discussion.  One is to show that we can
regard the chemical potential as the time component of a fictitious
gauge field of the $U(1)_B$ symmetry at zero temperature.  The other
is to find the criterion for the $U(1)_B$ {\it not}\/ to be
immediately broken in the presence of a chemical potential so we can
study the {\it dynamical}\/ breakdown.

The partition function in the grand canonical ensemble with nonzero
chemical potential can be calculated using the following path
integral: 
\begin{eqnarray} 
Z &=& \Tr\ e^{-\beta(H-\mu N)} \nonumber \\
&=& C \int \mathcal{D}\pi^{\dagger}\mathcal{D}\pi
\int \mathcal{D} \phi^{\dagger} \mathcal{D}\phi\
e^{\int_0^\beta d\tau \int \! d^3 x \,\left[
    i\pi \frac{\partial\phi}{\partial\tau} 
    +i\pi^{\dagger} \frac{\partial\phi^{\dagger}}{\partial\tau}
    -\mathcal{H}(\pi,\phi) +
    \mu \mathcal{N}(\pi,\phi) \right]} 
\end{eqnarray}
where $\mathcal{N}$ is the time-component of some conserved
current. We consider the case of a complex scalar field, where the
Hamiltonian is
\begin{equation}
\mathcal{H}=\pi^\dag \pi + \grad\phi^\dag \cdot
\grad\phi + m^2 \phi^\dag \phi
\end{equation}
and the conserved current is
$J_\mu = i(\phi^\dag \del_\mu \phi - \phi \del_\mu \phi^\dag)$. 
Thus $\mathcal{N}=i(\phi^\dag \pi^\dag - \phi\pi)$. The integrand of
the exponent in the path integral can be rewritten 
\bea
\lefteqn{ i\left(\pi^\dag \partial_\tau \phi^\dag+\pi
\partial_\tau \phi\right) -\left(\pi^\dag \pi +
\grad\phi^\dag \cdot 
\grad\phi + m^2 \phi^\dag \phi\right) + i\mu\left(\pi^\dag \phi^\dag -
\pi\phi \right) } \nonumber \\
&=& -\left( \pi^\dag -i(\partial_\tau -
\mu)\phi\right)\left(\pi-i(\partial_\tau + \mu)\phi^\dag\right) -
(\del_\tau+\mu)\phi^\dag(\del_\tau-\mu)\phi\nonumber \\ 
&& -\grad\phi^\dag \cdot \grad\phi - m^2 \phi^\dag \phi
\eea
Performing the functional integration over $\pi$ and $\pi^\dag$ leads
to the following expression for the partition function:
\beq
Z = C' \int \mathcal{D}\phi^\dag \mathcal{D} \phi\ e^{-\int_0^\beta
d\tau \int \! d^3 x \,\left[
(\del_\tau+\mu)\phi^\dag(\del_\tau-\mu)\phi + \grad\phi^\dag \cdot 
\grad\phi + m^2 \phi^\dag \phi \right]} \nonumber
\eeq
which can be analytically continued
to Minkowski space to yield an effective Lagrangian:
\begin{eqnarray}
&& C' \int
\mathcal{D}\phi^\dag \mathcal{D} \phi\ e^{i\int
dt \int \! d^3 x \,\left[
(\del_t+i\mu)\phi^\dag(\del_t-i\mu)\phi - \grad\phi^\dag \cdot 
\grad\phi - m^2 \phi^\dag \phi \right]} \nonumber \\ &&\equiv C' \int
\mathcal{D}\phi^\dag \mathcal{D} \phi\ e^{i\int 
dt \int \! d^3 x \,\mathcal{L}_\mathrm{eff} }
\end{eqnarray}

It is important to note that $\mathcal{L}_\mathrm{eff}$ is not simply
$\mathcal{L}+\mu\mathcal{N}$, since $\mathcal{N}$ is a function of
$\pi$ in addition to $\phi$. Instead,
\beq
\label{eqn:Lscalar}
\mathcal{L}_\mathrm{eff} = \partial^\nu \phi^\dag \partial_\nu \phi
+ i\mu \left(\phi^\dag \del_t \phi-\phi \del_t
\phi^\dag \right) - (m^2-\mu^2)
\phi^\dag \phi.
\eeq
The term linear in $\mu$ is the expected $\mu \mathcal{N}$
contribution. The term quadratic in $\mu$ arises from the modification
of the conjugate momenta $\pi=\dot{\phi}^\dag+i\mu \phi$. Here we see
immediately that for $m^2>0$ there is a critical chemical potential,
$|\mu_c|=m$ such that for $|\mu| > |\mu_c|$ the scalar potential is
unstable and Bose-Einstein condensation takes place. This can also be
seen explicitly by using the result for the transition temperature for
$\phi^4$ theory as in~\cite{Kapusta}:
\beq
\label{T_c}
T_c^2 = \frac{3}{\lambda} (\mu^2-m^2).
\eeq
This demonstrates again that condensation occurs for $\mu^2 > m^2$. 

In Appendix~\ref{HOexample} we test this method of adding a chemical
potential in quantum mechanics. We do this by comparing the partition
function computed directly with the one obtained using the effective
Lagrangian derived above.  The system we consider is a two-dimensional
harmonic oscillator with an added source term for $z$-component of
angular momentum which plays the role of the conserved current.

The purpose of the preceding exercise was to demonstrate that the
inclusion of a chemical potential in the Lagrangian amounts to
modifying the time derivative $\partial_t \to \partial_t -i\mu$, which
is equivalent to the addition of a non-dynamical gauge field that
acquires a nonzero vacuum expectation value (VEV) for its time
component.  Explicitly, the effective Lagrangian of
Equation~(\ref{eqn:Lscalar}) can be rewritten as
\beq
\mathcal{L}_\mathrm{eff} = (D^\mu \phi)^\dag D_\mu \phi - m^2
\phi^\dag \phi
\eeq
where
\beq
D_\mu \phi = (\del_\mu - ig A_\mu)\phi \
\ \ \mathrm{and} \ \ \ \langle A_\mu \rangle = \left(\frac{\mu}{g},
0, 0, 0 \right).
\eeq
It can be shown that the same ``covariant derivative'' gives the
correct effective Lagrangian for a theory including fermions and is
therefore suitable for use in supersymmetric theories. The advantage
of this formalism is that the coupling of the chemical potential to
the low energy degrees of freedom is determined by their $U(1)_B$
quantum numbers.

\section{SQCD with a Baryon Chemical Potential} \label{density}

We will study supersymmetric QCD which is defined to be an $SU(N_c)$
super Yang-Mills theory with $N_f$ flavors of chiral superfield quarks
$Q_i$ in the fundamental $N_c$ representation of the gauge group,
along with $N_f$ flavors of chiral ``anti-quarks'' $\Qbar_i$ in the
conjugate representation. This theory has the anomaly free global
symmetry $SU(N_f)_L\times SU(N_f)_R\times U(1)_B\times U(1)_R$. A
frequently appearing quantity is the one-loop beta function
coefficient $b_0 = 3N_c-N_f$. Much is already known about the infrared
limit of SQCD, which depends on the relative numbers of flavors and
colors. For $N_f\le N_c+1$ the low energy description is in terms of
composite mesons and baryons, whereas for higher $N_f$ it is in terms
of mesons and dual quarks.

In this paper we study SQCD with a baryon chemical potential. The
regular SQCD Lagrangian is defined at a high UV scale, where we also
add chemical potential terms associated with baryon number for both
fermions and bosons. In order to most easily apply the known exact
results for supersymmetric gauge theories mentioned
above~\cite{Intriligator:1995au}, we will follow the discussion in the
previous section and incorporate the added chemical potential as a
background (fictitious) $U(1)_B$ super-gauge field that has received
the appropriate vacuum expectation value.  The resulting UV Lagrangian
is
\begin{equation}
\label{Luv2}
\mathcal{L}_{UV}=\int d^4 \theta \left( Q_i^\dag e^{g_B V_B} Q_i
+\Qbar_i^\dag e^{-g_B V_B}\Qbar_i\right) + \frac{1}{g^2}\int d^2
\theta \left(\mathcal{W}_\alpha
\mathcal{W}^\alpha + \mbox{h.c.}\right)
\end{equation}
where
\begin{equation}
\label{background}
\langle V_B \rangle=\bar{\theta}\sigma^{\nu} \theta \langle A_\nu 
\rangle
\qquad \mbox{with} \qquad 
\langle{A_\nu} \rangle =\left(\frac{\mu}{g_B},0,0,0 \right).
\end{equation}
In Equation~(\ref{Luv2}) the $SU(N_c)$ gauge coupling has been
suppressed and the second term gives the gauge kinetic terms for the
$SU(N_c)$ gauge bosons but not for the $U(1)_B$, which remains
non-dynamical.

Note that the VEV in Equation~(\ref{background}) explicitly breaks
supersymmetry, so one may worry that this invalidates the powerful
supersymmetric results, namely that color is confined and that the low
energy degrees of freedom can be described by mesons, baryons, or dual
quark superfields. We will therefore consider a chemical potential
that is small compared to the dynamical scale $\Lambda$. In such a
case we can treat the chemical potential as a perturbation to the
supersymmetric dynamics.

As we have seen above, the chemical potential $\mu$ contributes a
tachyonic mass term to scalar potentials as in
Equation~(\ref{eqn:Lscalar}). For SQCD with massless quark
superfields, this means that squark condensation is immediately
favored. Because we are interested in studying the effect of strong
dynamics on condensation and symmetry breaking, we will add a
stabilizing mass for the squarks that returns the squark VEV to the
origin. From this stable UV theory we can move on to investigate the
symmetry breaking patterns that are triggered in the IR description of
the theory by strong dynamics.

We will consider both supersymmetric and SUSY-breaking masses to
stabilize the theory in the UV. A supersymmetric mass term appears in
the superpotential: \beq \label{eqn:Wmass} W_\mathrm{mass} = \sum_{ij}
m_{ij}Q_i\Qbar_j = \Tr (mM).  \eeq In order to preserve the most
global symmetry, we take the SUSY mass to have the form $m_{ij} =
m\delta_{ij}$, which explicitly breaks the chiral flavor symmetry to
the diagonal vector subgroup and leaves $U(1)_B$ intact. Stability
further requires $\mu < m$. Such supersymmetric mass terms are nice
because they do not damage the exact results of SQCD.

If one wishes to maintain the full global symmetry, the stabilizing
masses must break SUSY. A supersymmetry-breaking soft mass may be
added to the Lagrangian:
\beq \label{eqn:Lmass}
\mathcal{L}_\mathrm{mass} = -\tm^2 \left( Q^\dag Q + \Qbar^\dag \Qbar
\right),
\eeq
where the $Q$'s here represent the scalar components of the respective
superfields. Again, stability requires $\mu < \tm$. Since we rely on
many exact results that are only valid when the theory is nearly
supersymmetric, we will also require $\tm \ll \Lambda$.  However, the
presence of soft masses for the squarks in the UV theory may alter the
low energy potential for mesons and baryons (or dual quarks). This
question was addressed in~\cite{Arkani-Hamed:1998wc}
and~\cite{Luty:1999qc} using spurion arguments (and from a different
approach in \cite{Cheng:1998xg}).  The main observation is that scalar
soft-masses can be incorporated in a superfield $\mathcal{Z}$ whose
couplings are determined by an anomalous $U(1)_A$ symmetry and
RG-invariance. Results for softly broken SUSY are reviewed in the
following sections and in greater detail in
Appendix~\ref{exactresults}.

We are now ready to investigate the effects of adding stabilizing
masses and a chemical potential to a UV SQCD theory.  In each of the
theories below we will first determine the symmetry pattern in the
absence of a chemical potential, and then check whether adding a
chemical potential breaks symmetries. We are be particularly
interested in comparing $U(1)_B$ and its breaking pattern to the
non-supersymmetric case.  Since the IR degrees of freedom depend on
the number of flavors, we will consider separate cases in the
following order: $N_f = N_c$ (quantum modified moduli space), $N_f <
N_c$ (Affleck--Dine--Seiberg superpotential), $N_f = N_c+1$
(confinement without chiral symmetry breaking), and $N_c + 2 \le N_f <
3 N_c$ (dual picture).

\section{$N_f=N_c$}

When the number of flavors and colors are equal the infrared
description is in terms of gauge invariant chiral superfields called
``mesons'' $M_{ij}=Q^a_i\Qbar^a_j$ and ``baryons''
$B=\epsilon_{i_1\cdots i_{N_f}} Q_{i_1}\cdots Q_{i_{N_f}}$ and
$\Bbar=\epsilon_{i_1\cdots i_{N_f}} \Qbar_{i_1}\cdots
\Qbar_{i_{N_f}}$. Here the suppressed color indices are completely
antisymmetrized. The mesons transform as $(N_f, \Nbar_f)$ under
the flavor symmetries, whereas the baryons are flavor singlets. The
meson and baryon superfields satisfy a constraint that is modified
quantum mechanically and can be implemented in a dynamical
superpotential with a Lagrange multiplier superfield $X$:
\beq \label{eqn:Wconstraint}
W=\frac{X}{\Lambda^{2N_c-2}} 
\left( \mathrm{det} M -B\Bbar - \Lambda^{2N_c} \right),
\eeq
where $\Lambda$ is the scale of the $SU(N_c)$ theory. The equation of
motion for $X$ enforces the constraint.

We will now add stabilizing masses. In the following subsection we
will add supersymmetric masses. In Section~\ref{SUSYbreak} we will add
soft masses and discover that we need to choose a vacuum to expand
around. In \ref{chiralbreak} and \ref{baryonbreak} we will consider
two such vacua in turn.

\subsection{Supersymmetric Mass} \label{SUSYstable}

As discussed above, the theory with massless squarks in the UV is
immediately unstable upon the introduction of a chemical potential. We
will first consider supersymmetric mass to stabilize the theory in the
UV. When the supersymmetric mass term in Equation~(\ref{eqn:Wmass}) is
added to the superpotential (\ref{eqn:Wconstraint}) the $F$-term
equations force the baryons to vanish, but yield the VEVs \beq \langle
M_{ij}\rangle = (m^{-1})_{ij}\left( \Lambda^{b_0} \det m
\right)^{1/N_c} \qquad \mathrm{and}\qquad \langle X^{N_c}\rangle =
-\frac{\det m}{(\det M)^{N_c-1}}.  \eeq The VEV for $X$ then manifests
itself as a mass term for the baryons.
The low energy Lagrangian for the baryons is thus
\beq
\mathcal{L} = \int d^2\theta d^2\bar{\theta}
\frac{c_0}{\Lambda^{2N_c-2}}(B^\dag e^{N_cV_B} B + \Bbar^\dag
e^{-N_cV_B}\Bbar) + 
\int d^2\theta \frac{m}{\Lambda^{N_c-1}} \Bbar B
\eeq
where $c_0$ is an unknown coefficient in the K\"ahler potential which
is estimated in naive dimensional analysis to be $\mathcal{O}(1)$.
We've shown that at zero chemical potential $U(1)_B$ is conserved and
the chiral symmetry is broken to its vector subgroup.

We will now show that this changes as a chemical potential is added.
After canonically normalizing $B$ and $\Bbar$ and inserting the VEV
of $V_B$ from Equation~(\ref{background}) to incorporate the chemical
potential, the squared mass of the
baryons is
\beq
m^2_{B,\Bbar}=m^2/c_0-N_c^2\mu^2,
\eeq
 which means there is a critical chemical potential
\beq
\mu_c^2 = \frac{m^2}{c_0 N_c^2},
\eeq
such that for $\mu >\mu_c$ the baryon mass-squared becomes negative,
drawing the baryon VEVs away from the origin. We cannot determine
exactly where the baryons stabilize, but due to the presence of the
SUSY masses we know that the UV theory is stable, which means the
baryon VEVs should be of order $\Lambda$. Thus we conclude that as
long as $c_0 N_c^2 > 1$ there are values of $\mu$ such that $\mu_c <
\mu < m$ where our approximations are valid and the $U(1)_B$ symmetry
is spontaneously broken. This is depicted in the left diagram of
Figure~\ref{fig:mum-plot}.

\subsection{SUSY-Breaking Mass} \label{SUSYbreak}

Instead of adding supersymmetric mass terms, we now consider the
addition of soft SUSY-breaking mass terms to stabilize the UV
theory. In order to determine the potential for the low energy degrees
of freedom due to SUSY breaking, we will adopt a dimensionless
convention for the fields~\cite{Luty:1999qc}
\begin{equation}
\Mh \equiv \frac{M}{\Lambda_h^2} \qquad
\Bh \equiv \frac{B}{\Lambda_h^{N_c}} \qquad
\Bbarh \equiv \frac{\Bbar}{\Lambda_h^{N_c}}.
\end{equation}
The constraint then takes the convenient form
\begin{equation}
\label{constraint}
\det \Mh - \Bh \Bbarh =1.
\end{equation}
The fields $\Mh$, $\Bh$ and $\Bbarh$ are dimensionless and are
uncharged under both the anomalous $U(1)_A$ symmetry described in
Appendix \ref{exactresults} and the anomaly free $U(1)_R$. This means
we cannot use these symmetries to constrain the theory and cannot
determine the IR soft masses exactly.  However, we will still be able
to maintain some control over the sign of the baryon masses and thus
rule out some symmetry preserving vacua due to their instability.

The flavor and $U(1)_B$ symmetries constrain the
K\"{a}hler potential to be a some function $K$ of real and invariant
combinations of the fields
\beq
\label{kahlerN}
\mathcal{L}\supset \int d^2\theta d^2\bar{\theta}\left[ I \times
K \left(\Bh^\dag e^{N_cV_B} \Bh,\ 
\Bbarh ^\dag e^{-N_cV_B}\Bbarh,\ 
(\Bh\Bbarh + \mbox{h.c.} ),\ 
\mathrm{tr} \Mh^\dag \Mh,\ \ldots \right)\right]
\eeq
where we've only written lowest order terms in the fields. Here $I$ is
a $U(1)_A$ and RG-invariant superfield defined in
Equation~(\ref{eqn:Idef}). The point of Equation~(\ref{kahlerN}) is
that $I$, which contains the soft mass $\tm^2$ as its
$\theta^2\bar{\theta}^2$ component, only appears as an overall
multiplicative factor, while the background $U(1)_B$ gauge field which
contains the chemical potential $\mu$ couples to the baryons as
expected. Note that the $\Bbarh \Bh$ term cannot be forbidden by the
symmetries, and prevents us from completely determining the baryon
soft masses. In later cases such terms will be disallowed by $U(1)_A$
invariance.

In what follows we will assume that the SUSY-breaking effects are
sufficiently small compared to the dynamical scale $\Lambda$ such that
the quantum modified constraint, Equation~(\ref{constraint}), is
strictly enforced. We will therefore restrict ourselves to
the supersymmetric flat directions while enforcing the constraint by
hand. We then minimize the resulting potential including the
soft-breaking terms.

Since the origin of moduli space is excluded by
the constraint, some of the global symmetries are
necessarily broken by the strong dynamics even in the absence of a
chemical potential.  We will therefore expand the low energy
K\"{a}hler potential around points that possess the highest degree of
remaining symmetry and see whether they remain stable in the presence
of a chemical potential. In the following subsections we will analyze
the two points of ``maximal'' symmetry.
We cannot determine which point gives the true vacuum of softly-broken
supersymmetric gauge theory.  However, the
first point resembles the dynamics of non-supersymmetric theory the
most and hence is a good candidate for the true vacuum.  In that case we 
can show that a large enough chemical potential destabilizes the vacuum,
leading to the spontaneous breakdown of $U(1)_B$.

\subsubsection{Vacuum with Chiral Symmetry Breaking} \label{chiralbreak}

In terms of the hatted fields, a general vacuum that conserves baryon
number has $\langle \Bh \rangle=\langle \Bbarh \rangle=0$ and $\langle
\det \Mh \rangle=1$.  Among them, the solution to the constraint
$\langle \Mh_{ij} \rangle=\delta_{ij}$ preserves the largest symmetry.
At this vacuum of the supersymmetric theory, the $SU(N_f) \times
SU(N_f)$ chiral symmetry is broken to the diagonal $SU(N_f)$, while
$U(1)_B$ is unbroken.  Therefore this point resembles strongly the
dynamics of the non-supersymmetric theory, and hence it is reasonable
to expect that the soft supersymmetry breaking effects make this point
stable by giving the fluctuations positive squared masses. Indeed, we
will show below that this is not unreasonable. Furthermore, since this
pattern is similar to that achieved by adding supersymmetric masses it
is interesting to compare the results for the two types of masses once
a chemical potential is added.

We expand the meson field as
$\Mh=\delta_{ij}+\Phi_{ij}$. Following~\cite{Luty:1999qc}, we can now
use Equation~(\ref{constraint}) to eliminate one of the degrees of
freedom. It is convenient to express the trace of $\Phi$ in terms of
the other fields,
\begin{equation}
\label{newconstraint}
\tr\Phi=\Bbarh\Bh+\frac{1}{2} \tr( \Phi'^2 )+ 
\mbox{higher terms in}\ \tr\Phi,
\end{equation}
where $\Phi'_{ij}=\Phi_{ij}-\delta_{ij}/N_f$ is the traceless
component of $\Phi$.

The K\"{a}hler potential in Equation~(\ref{kahlerN}) can now be
expressed in terms of the small excitations $\Phi',\Bh$ and $\Bbarh$
and expanded in a power series. For example, substituting into the
meson kinetic term in the K\"ahler potential we get
\begin{eqnarray}
\label{mdagm}
\tr\Mh^\dag\Mh &=& N_f+\tr(\Phi'^\dag \Phi')+ \frac{1}{2}
\left(\tr(\Phi'^2)
+\mbox{h.c.}\right)+\left(\Bbar\Bh +\mbox{h.c.} \right) +\ldots
\nonumber\\ &=& N_f+ \frac{1}{2} \tr\left( \Phi'+\Phi'^\dag \right)^2
+\left(\Bbar\Bh +\mbox{h.c.} \right) +\ldots
\end{eqnarray}

The soft masses will ultimately be determined by inserting the SUSY
breaking VEVs for $V_B$ and $I$ (Equations~(\ref{background}) and
(\ref{Icomponents} respectively) and performing the $d^4\theta$
integral in the K\"ahler potential. The mesons will receive a soft
mass only from the SUSY breaking in $I$. From Equation~(\ref{mdagm})
we see that at lowest order in the fields only the real part of
$\Phi'$ receives a mass while the imaginary part remains
massless. However, since we are expanding the meson field around a VEV
of order the dynamical scale, masses that come from inserting the VEV
into the higher order terms in Equation~(\ref{kahlerN}) will not be
suppressed.

We can use Goldstone's Theorem to show that the imaginary part of
$\Phi'$ remains massless
to all orders in the fields.  The K\"{a}hler potential of 
Equation~(\ref{kahlerN}) can be expanded in powers of the
excitations\footnote{We are assuming that the strong dynamics respect
a $Z_2$ symmetry that interchanges $B$ and $\Bbar$.}
\begin{eqnarray}
\mathcal{L}&\supset& \int d^2 \theta d^2 \bar{\theta}\ I \times
\left[ c_1\left(\Bh^\dag e^{N_cV_B} \Bh +\Bbarh ^\dag
e^{-N_cV_B}\Bbarh\right)\right. \nonumber\\
&+& \left(c_2\Bh\Bbarh+\mbox{h.c.}\right)+c_3\tr\Phi'^\dag\Phi'
+c_4\left(\tr\Phi'^2+\mbox{h.c.}\right)\left. \right].
\end{eqnarray}
The coefficients $c_i$ are derivatives of the function $K$ and include
contributions from operators of all orders in the meson field. The
coefficients $c_1$ and $c_3$ should be absorbed into the definition of
the fields in order to have canonically normalized kinetic terms.
Performing the superspace integration, the mass matrix for each
$\Phi'_{ij}$ in the basis $(\Phi', \Phi'^\dag)$ is
\begin{equation}
\label{m_phi}
m^2_{\Phi'}= \tm^2
\left( \begin{array}{cc} 
1 & c_4/c_3 \\ 
 c_4/c_3 & 1
\end{array}\right).
\end{equation}
Since the VEV of $M$ breaks the chiral flavor symmetry $SU(N)_L\times
SU(N)_R$ to the diagonal $SU(N)_{L+R}$ there must be $N_f^2-1$
massless Goldsone Bosons. Since there are $N_f^2-1$ $\Phi'$ fields
this can only work out if all the matrices in Equation~(\ref{m_phi})
have a zero eigenvalue. This sets the ratio $c_4/c_3=1$ (though -1
will do as well) and the masses are
\begin{equation}
\label{mPhi}
m^2_{\mbox{Re}\Phi'}=2\tm^2 \qquad
m^2_{\mbox{Im}\Phi'}=0
\end{equation}
Therefore, the fluctuations in the meson degrees freedom indeed have
positive mass-squared for their real parts and zero mass for the
Nambu--Goldstone bosons, and hence the vacuum is stable against these
fluctuations. Since the mesons are uncharged under baryon number this
is unchanged in the presence of a chemical potential.

On the other hand, the situation with the baryonic degrees freedom is
less clear.  In~\cite{Luty:1999qc} the meson masses were determined as
above and it was noted that since we do not have control over the
coefficients of the $\Bbar B$-type mass, the baryon masses cannot be
explicitly determined. Nonetheless, we can still learn about the
stability of the potential. After canonically normalizing and
performing the $\theta^4$ integral, the baryon mass matrix in the
basis $(\Bh\
\Bbarh^\dag)$ is 
\beq m_B^2 =\left( \begin{array}{cc}
    \tm^2-N_c^2 \mu^2 & \frac{c_2}{c_1}\tm^2 \\
    \frac{c_2^*}{c_1}\tm^2 & \tm^2-N_c^2 \mu^2
\end{array}\right) \label{eq:Bmatrix}
\eeq
The diagonal mass receives a positive contribution from the
stabilizing soft mass $\tm^2$ and a negative contribution from the
chemical potential which is enhanced by a factor of $N_c^2$. Recall
that the requirement for a stable theory is $\tm^2>\mu^2$.

In order to determine whether a chemical potential induces a phase
transition we first must check whether baryon number is conserved at
$\mu=0$.  In the absence of the chemical potential, the eigenvalues of
the mass matrix Eq.~(\ref{eq:Bmatrix}) are $(1 \pm
\left|\frac{c_2}{c_1}\right|) \tm^2$. Therefore the vacuum $B=\Bbar=0$
is stable if the off-diagonal element is smaller than the diagonal
element, $|c_2| < |c_1|$. We cannot rigorously justify that this is
the case, though it is a reasonable assumption.\footnote{For $N_c=2$
the $SU(2)\times SU(2)$ flavor symmetry is enlarged to $SO(4)$, which
prohibits the $B\Bbar$ terms. However, the $SO(4)$ is broken by the
chemical potential which would be expected to generate a $B\Bbar$ term
with a coefficient of order $\mu/\Lambda$, which is less than the
$\mathcal{O}(1)$ coefficient of the $B^\dag B$ terms. Thus
$|c_2|<|c_1|$ is the natural expectation for $N_c=2$.} Henceforth we
assume that this is the case.  Then the vacuum is stable against
baryonic fluctuations and the dynamics is precisely that of the
non-supersymmetric theory: dynamical chiral symmetry breaking with
massless Nambu--Goldstone bosons and massive baryons.  The main
difference, however, is that the baryons are not as heavy as $\Lambda$
but rather as light as $\tm \ll \Lambda$ (within the validity of our
approximations).

Now we consider a finite chemical potential.  As seen from the mass
matrix Eq.~(\ref{eq:Bmatrix}), there is a critical value of the
chemical potential that makes one of the eigenvalues negative:
\begin{equation}
  \mu_c^2 = \frac{1}{N_c^2} \left(1 - \left|\frac{c_2}{c_1}\right|
  \right) \tm^2.
\end{equation}
Without loss of generality, we can always perform a $U(1)_B$ rotation
to make $c_2/c_1 < 0$.  Then the direction of the instability is $\Bh
= \Bbarh$.  The fields roll down the potential and dynamically break
the $U(1)_B$ symmetry.  On the other hand, $|\Bh|, |\Bbarh| \gg 1$
corresponds to the semi-classical regime where
\begin{eqnarray}
  Q = \Qbar = \left( 
    \begin{array}{ccc}
      v & &\\
      & \ddots &\\
      & & v
    \end{array} \right), \qquad
  M = \left(
    \begin{array}{ccc}
      v^2 & &\\
      & \ddots &\\
      & & v^2
    \end{array} \right), \qquad
  B = \Bbar = v^{N_c}.
\end{eqnarray}
We know the potential is stable in this regime because the dynamics is
correctly described in terms of the quark degrees of freedom and $\tm
> \mu$ assures the stability.  Therefore, the baryon fields should
stabilize somewhere around $\Bh = \Bbarh \sim 1$.  The precise vacuum
expectation values depend on the exact form of the K\"ahler potential
and hence cannot be worked out.  However the existence of a stable
vacuum is obvious from this discussion.

In summary, up to the assumption of $|c_2| < |c_1|$, the
supersymmetric vacuum with chiral symmetry breaking and unbroken
$U(1)_B$ is stable in the presence of the soft supersymmetry
breaking.\footnote{If $|c_2| > |c_1|$, baryon number is already broken
in the absence of a chemical potential. In this case a chemical
potential does not change the dynamics qualitatively but it does cause
the $U(1)_B$-breaking to be ``stronger''. This will cause the critical
temperature at which $U(1)_B$ is restored to grow with $\mu$ as one
would naively expect (see Equation~\ref{T_c}).}  There is a critical
chemical potential of the order of supersymmetry breaking beyond which
$U(1)_B$ breaks dynamically.  The order parameters are expected to all
be on the order of the dynamical scale and hence the theory is
strongly coupled.  Nonetheless the breakdown of $U(1)_B$ and the
presence of a stable vacuum is guaranteed.

This same conclusion holds even if we were to expand around more
general baryon number conserving vacua, since the crucial piece of
information we need is only the instability of the baryon directions,
which follows directly from the diagonal soft mass and chemical
potential contributions to the baryon mass terms. Since this is
independent of the precise meson VEV, we conclude that (modulo one
assumption) \emph{all} baryon conserving vacua are unstable and thus
that baryon number is not a global symmetry in the presence of a
chemical potential that is larger than $\mu_c$. As in the case with
the supersymmetric mass, the critical chemical potential is again of
order $m/N_c$. The breaking pattern also agrees with the result for
non-supersymmetric QCD.

\subsubsection{Vacuum with Baryon Number Breaking} \label{baryonbreak}

In the supersymmetric limit, there is another vacuum with a large
unbroken symmetry, $\langle\Mh\rangle=0$ and
$\langle\Bbarh\Bh\rangle=-1$. It preserves the full $SU(N_f) \times
SU(N_f)$ chiral symmetry but breaks baryon number symmetry.  We
already see that this point is of lesser interest to us since we are
focusing on the $U(1)_B$-breaking signature induced by the chemical
potential.

Again we would like to determine whether this point is stable in the
presence of soft supersymmetry breaking.  The dynamical degrees of
freedom around this point are the full meson matrix $M$ and another
chiral superfield $b$: $B = e^b$, $\Bbar = e^{-b}$. Since $b
\rightarrow b + i \theta$ under a $U(1)_B$ transformation, the
K\"ahler potential depends only on $b + b^*$.  In the absence of a
chemical potential, the K\"ahler potential for mesons takes the form
\begin{equation}
\mathcal{L}\supset\int d^4 \theta I f(b +b^*) {\rm Tr} M^\dagger M.
\end{equation}
The positivity of the kinetic term for
meson fields requires $f(0)>0$, while the $[\ln I]_{\theta^2
  \bar{\theta}^2} = - 2\frac{N_f}{b_0} \tm^2$ gives the mesons a
positive mass squared. Therefore, the meson directions are stable
around this point.

Concerning the direction $b$, the leading term in the K\"ahler
potential is \begin{equation}
\mathcal{L}\supset\int d^4 \theta \frac{1}{2}I (b+b^*)^2,
\end{equation}
which gives mass squared $\tm^2$ to the real part while
keeping the imaginary part a massless Nambu--Goldstone boson. A linear
term $\int d^4 \theta I (b+b^*)$ is forbidden by charge conjugation
invariance $Q \leftrightarrow \Qbar$ that requires the invariance of
the effective Lagrangian under $b \rightarrow -b$.  Therefore this
point is stable even in the presence of the soft supersymmetry
breaking.

Because baryon number is already broken in the absence of the chemical
potential, the finite chemical potential can change the physics
qualitatively only by breaking the chiral symmetry. However, it is the
breaking of baryon number that prevents us from calculating whether
this is indeed the case. One possibility is that the meson direction
is destabilized by an operator of the type
\begin{equation}
\mathcal{L}\supset
\int d\theta^2 d\bar{\theta}^2\ 
c I\left(\Bh^\dag e^{N_c V_B} \Bh\right)\left( \tr \Mh^\dag \Mh \right)
\supset c N_c^2 \mu^2 |B_0|^2 \tr \Mh^\dag \Mh
\end{equation}
which could give an additional contribution of either sign to the
meson mass squared. Thus we cannot determine the fate of the chiral
symmetry. In any case, this vacuum is of limited interest to us.

\section{$N_f < N_c$}

When $N_f<N_c$ the non-supersymmetric results leave an unbroken global
$U(1)$. As we will show below this is true in SQCD as well. In this
case the IR degrees of freedom consist only of the meson superfield
$M=Q\Qbar$ since there are not enough quark flavors to form
color-singlet baryons. The SQCD dynamics generates the so-called
Affleck--Dine--Seiberg superpotential,
\begin{equation}
  W = (N_c - N_f) \frac{\Lambda^{3+2N_f/(N_c-N_f)}}{({\rm
  det}M)^{1/(N_c-N_f)}}.
\end{equation}
Because of this potential, the meson field is driven away from the
origin, leading to a run-away theory with no vacuum.

However, this runaway behavior is stopped by the masses we add in the
UV.  Let us begin with a supersymmetric mass and assume that this mass
is large, $m>\Lambda$.  Adding this mass to the superpotential and
solving for the meson VEV we get the same formula as in the previous
section
\beq
\langle M_{ij}\rangle = (m^{-1})_{ij}\left( \Lambda^{b_0} \det m
\right)^{1-N_f/N_c}\sim\left( \frac{\Lambda}{m}\right)^{1/N_c}\Lambda^2.
\eeq
Note that the VEV is smaller than $\Lambda$ and therefore the meson is
indeed the appropriate degree of freedom to describe the IR vacuum. We
can immediately see that baryon number is unbroken because the IR
degrees of freedom are not charged under $U(1)_B$. For this case
adding a chemical potential has no effect.

We will now consider adding small masses $m<\Lambda$. In this case the
supersymmetric masses and the soft masses yield similar results so we
will treat them together. When the stabilizing mass is small the
mesons runs far away from the origin before it is stabilized at VEVs
much larger than $\Lambda$. This region of moduli space is described
in terms of quarks.  The large meson amplitude corresponds to the flat
direction

\begin{equation}
  Q = \Qbar = \left( 
    \begin{array}{ccc}
      v_1 & \cdots & 0 \\
      \vdots & \ddots & \vdots \\
      0 & \cdots & v_{N_f} \\
      \vdots &  & \vdots \\
      0 & \cdots & 0
    \end{array} \right)
\end{equation}

Rewriting the ADS potential in terms of quarks and adding soft or
supersymmetric masses and the tachyonic mass due to the chemical
potential we can minimize and solve for $v_1,\ldots,v_{N_F}$.  For
example, when soft masses are added such that $0 < \tm^2 -
\mu^2$, we get
\begin{equation}
  v_1 = \cdots = v_{N_f} \sim \Lambda 
  \left( \frac{\Lambda}{\sqrt{\tm^2-\mu^2}} \right)^{(N_c-N_f)/2N_c}.
\end{equation}
The form of the VEV is slightly different when a supersymmetric mass
is used, but $v_1=\ldots=v_{N_f}$ is still satisfied. Therefore this
theory dynamically breaks the chiral symmetry $SU(N_f)\times SU(N_f)
\rightarrow SU(N_f)$.  However, a $U(1)_{\widetilde{B}}$ remains
unbroken. The new baryon symmetry is the simultaneous transformation
under the original $U(1)_B$ and a $U(1)'$, where $U(1)'$ is a subgroup
of the gauge symmetry embedded as $SU(N_c)
\supset SU(N_f) \times SU(N_c-N_f) \times U(1)'$. This situation is
analogous to the non-supersymmetric case of $N_f = 2$ and $N_c =3$
(the so--called 2SC phase) where the unbroken $U(1)$ is a combination
of baryon number and a broken color generator.

\section{$N_f = N_c+1$} 

When $N_f=N_c+1$ the IR theory is again described in terms of mesons
$M_{ij}=Q^a_i\Qbar^a_j$ and baryons $B_i=\epsilon_{ij_1\cdots j_{N_c}}
Q_{j_1}\cdots Q_{j_{N_c}}$ and $\Bbar_i=\epsilon_{ij_1\cdots j_{N_c}}
\Qbar_{j_1}\cdots \Qbar_{j_{N_c}}$, but the symmetry transformation
properties are different. Choosing the quarks to have baryon number 1,
the fields $M$, $B$ and $\Bbar$ transform as
$(N_f,\Nbar_f)_0$, $(\Nbar_f,1)_{N_c}$ and
$(1,N_f)_{-N_c}$ respectively, under the global symmetry $SU(N_f)\times
SU(N_f)\times U(1)_B$. 

We summarize our results before presenting the details. Recall that
for $N_f=4$ in non-supersymmetric QCD baryon number is broken in the
presence of a high chemical potential. For SQCD, adding a
supersymmetric mass in a specified range will again lead to a critical
chemical potential above which baryon number is broken, in agreement
with the QCD result. Adding UV soft masses yields only partial
agreement.  If we choose to stay along a flat direction that breaks
the chiral symmetry to a vector symmetry we indeed find that the
existence of a critical chemical potential is plausible. However,
there is also another flat direction along which the chiral symmetry
is unbroken which has tachyonic masses for the baryons, thus breaking
$U(1)_B$ even for $\mu=0$.

\subsection{Supersymmetric Mass}

As in the previous case we will stabilize the UV theory first with
supersymmetric masses $m$ and then with soft SUSY-breaking masses
$\tm$.

The low energy superpotential for baryons and mesons, including the
supersymmetric quark masses, is
\beq \label{eqn:Wncplus1}
W=\frac{1}{\Lambda^{b_0}}(\mathrm{det} M - B_i M_{ij} \Bbar_j) + \Tr (mM).
\eeq
The baryon VEVs vanish but the mesons acquire a nonzero VEV:
\beq
\langle M_{ij} \rangle = (m^{-1})_{ij} \left(\Lambda^{b_0} \det m
\right)^{1/N_c}.
\eeq
The meson VEV yields a mass for the baryons $\int d^2\theta (\langle M
\rangle/\Lambda^{b_0}) \Bbar B$. Including the chemical potential
contribution, the mass-squared for the canonically normalized baryon
and antibaryon fields is
\beq
m^2_{B,\Bbar} =
\frac{1}{c_0}\left(\frac{m}{\Lambda}\right)^{2/N_c}\Lambda^2 -
N_c^2\mu^2
\eeq
where again $c_0$ is an unknown $\mathcal{O}(1)$ coefficient in the
K\"ahler potential, and we have taken the supersymmetric mass
$m_{ij}=m\delta_{ij}$ in order to preserve as much global symmetry as
possible.

Here again there is a critical chemical potential $\mu_c$ above which
the baryons become unstable and break $U(1)_B$:
\beq
\mu_c^2 = \frac{1}{c_0 N_c^2}\left(\frac{m}{\Lambda} \right)^{2/N_c}
\Lambda^2.
\eeq
However, this formula will only be valid as long as $\mu_c^2 <
\Lambda^2$ (near-SUSY limit) and $\mu_c^2 < m^2$ (UV stability).
These two constraints yield the requirements
\beq
(c_0 N_c^2)^{N_c/(2-2N_c)} < \frac{m}{\Lambda} < (c_0 N_c^2)^{N_c/2}.
\eeq
These can be satisfied as long as $c_0 N_c^2 >1$, since (for $N_c>1$)
the lower bound is a decreasing function of $c_0 N_c^2$ and the upper
bound is an increasing function. This is the same restriction on $c_0$
that we found in the case where $N_f=N_c$. We again conclude that it
is very likely that baryon number is broken by a sufficiently large
chemical potential.  There are, however, some differences compared to
the $N_f=N_c$ result.  The critical chemical potential is not
$\mathcal{O}(m)$ but is a combination of $m$ and $\Lambda$. Also note
that the UV mass is bounded from below in order for our result to
hold, as shown in the right diagram of Figure~\ref{fig:mum-plot}.

\subsection{SUSY-Breaking Mass}

Now we consider a different situation where $m$ is set equal to zero
in Equation~(\ref{eqn:Wncplus1}) and instead we add universal soft
SUSY-breaking masses $\tm$ for the squarks in the UV theory. Using the
results of~\cite{Arkani-Hamed:1998wc} reviewed in Appendix
\ref{exactresults}, the soft-masses for the squarks result in the
following soft masses for the mesons and baryons:
\beq
\label{softmasses}
\tm^2_M = \frac{2N_c-4}{2N_c-1} \tm^2 
\ \ \ \ \mathrm{and} \ \ \ \
\tm^2_{B,\Bbar} = \frac{2-N_c}{2N_c-1} \tm^2 - N_c^2 \mu^2.
\eeq
Notice that for $N_c \geq 3$ the meson mass-squared is positive, but
the baryons are tachyonic. Here we have also included the contribution
from the chemical potential, which simply follows from baryon number.

When we canonically normalize the baryon and meson superfields the
unknown constants $c_{M,B,\Bbar}$ reappear as relative coefficients
between the terms of the superpotential:
\beq
W=c_1 \frac{\mathrm{det} M'}{\Lambda_h^{N_f-3}} - c_2 B'_i M'_{ij}
\Bbar'_j
\eeq
where $M'=M/\Lambda_h$ and $B'=B/\Lambda_h^{N_c-1}$ are dimension one
fields, and $c_1=c_M^{-N_f/2}$ and $c_2=(c_M c_B
c_\Bbar)^{-1/2}$. This, along with the soft masses of
Equation~(\ref{softmasses}), gives rise to the following scalar
potential
\begin{eqnarray}
\label{Vir}
V_{IR} &=&  \left|
c_1 \frac{\tm'_{ij}}{\Lambda_h^{N_f-3}}-c_2 B'_i \Bbar'_j
\right|^2 
+ \left| c_2 M'_{ij} \Bbar'_j \right|^2 
+\left| c_2 B'_i M'_{ij} \right|^2 \nonumber\\
&&+\sum_{ij}\tm^2_M \left| M'_{ij} \right|^2
+\sum_{i}\tm^2_B \left| B'_{i} \right|^2
+\sum_{i}\tm^2_{\Bbar} \left| \Bbar'_{i} \right|^2
\end{eqnarray}
where $\tm'_{ij}$ is the $ij$th cofactor of $M'_{ij}$. In the
supersymmetric limit this potential has two flat direction with
different symmetry breaking patterns. We will now analyze these flat
directions separately.

The first flat direction has VEVs of the form:
\beq 
\label{eqn:flatdir}
B=\Bbar = \left(\begin{array}{c} b \\ 0 \\ \vdots \\ 0 \end{array}
\right) \ \ \ \ \ \ \ \ M=\left(\begin{array}{cccc} 0 & & & \\ & a & &
\\ & & a & \\ & & & \ddots \end{array} \right)
\eeq
Along this flat direction the relationship between the VEVs is
constrained to be $c_1 a^{N_c} = c_2 \Lambda^{N_c-2} b^2$ by the first
term in Equation~(\ref{Vir}). The introduction of the soft masses has
the potential to destabilize this flat direction. The potential along
the flat direction for $|a| < \Lambda$ is given by
\begin{equation}
\label{Virsmallm}
  V_{IR} = N_c \frac{2N_c-4}{2N_c-1} \tm^2 |a|^2
  - 2 \left( \frac{N_c-2}{2N_c-1} \tm^2 + N_c^2 \mu^2 \right) \left|
    \frac{c_1}{c_2} \frac{a^{N_c}}{\Lambda^{N_c-2}} \right|.
\end{equation}
On the other hand, at large amplitudes the quark description is
valid.  The flat direction above corresponds to the quark VEVs
\begin{equation}
  Q = \Qbar = \left( 
    \begin{array}{cccc|c}
      0 & & & & 0\\
      & \sqrt{a} & & & 0\\
      & & \ddots & & \vdots\\
      & & & \sqrt{a} & 0
    \end{array} \right)
\end{equation}
and hence, for $|a|>\Lambda$,
\begin{equation}
\label{Virlargem}
  V_{IR} = 2 N_c (\tm^2 - \mu^2) |a|.
\end{equation}
Clearly $V_{IR}=0$ at the origin. The potential goes up as $|a|^{2}$
first, but then goes down as $-|a|^{N_c}$ at a larger amplitude. We
will have a vacuum with energy lower than the origin if the potential
in Equation~(\ref{Virsmallm}) becomes negative within its range of
applicability, namely $|a| < \Lambda$.  Beyond this region the
potential goes up linearly with $|a|$ as in Eq. (\ref{Virlargem}).

When $\mu=0$, $V_{IR}$ will be negative when
\beq
\left( N_c \left|\frac{c_2}{c_1}\right| \right)^{1/(N_c-2)} <
\frac{|a|}{\Lambda} < 1
\eeq
where the upper-bound is the requirement for
Equation~(\ref{Virsmallm}) to be valid. This condition is somewhat
restrictive, though by no means impossible to satisfy. For $N_c=3$ it
requires $|c_1|>N_c |c_2|$ with $|a|$ near its upper bound.

Nonzero $\mu$ helps make $V_{IR}$ more negative. In particular, there
is a critical chemical potential given by
\beq
\mu_c^2 = \frac{N_c-2}{N_c^2(2N_c-1)}\left[ N_c
\left|\frac{c_2}{c_1}\right| \left(\frac{\Lambda}{|a|}
\right)^{N_c-2}-1  \right] \tm^2
\eeq
above which $V_{IR}$ goes negative, as long as
\beq
\left(\left|\frac{c_2}{c_1}  \right|
\left[\frac{N_c(N_c-2)}{N_c^2(2N_c-1)+(N_c-2)}  \right]
\right)^{1/(N_c-2)} < \frac{|a|}{\Lambda} < 1.
\eeq
The lower bound comes from requiring $\mu<\tm$, whereas the upper
bound is again required for Equation~(\ref{Virsmallm}) to be
valid. The extra factors of $N_c$ on the left-hand side make the
requirements on $c_1$ and $c_2$ less stringent. However, as $N_c$
grows the allowed range for $|a|$ decreases. If indeed such a point is
the global minimum, baryon number will be broken, and the chiral
flavor symmetries will be broken to the diagonal $SU(N_f-1)_V$ vector
symmetry once $\mu$ exceeds $\mu_c$.

We will now analyze the second flat direction.  When $M=0$ a SUSY
vacuum requires at least one of $B_i$ or $\Bbar_i$ to also vanish. For
example, take $B_i=0$ for all $i$. The positive meson mass-squared
will tend to keep the meson VEV zero, but the negative baryon
mass-squared will draw the $\Bbar$ VEVs away from the
origin. Specifically, flavor symmetry allows us to write the VEVs as
\beq
B= \left(\begin{array}{c} b \\ 0 \\ \vdots \\ 0 \end{array}\right)
\ \ \ \mathrm{and} \ \ \ \Bbar = \left(\begin{array}{c} \bbar \\ 0 \\
\vdots \\ 0 \end{array}\right)
\eeq
which will be pulled in the direction $|b|^2 \neq 0, |\bbar|^2=0$ or
vice versa. In the former case the flavor symmetry is broken to
$SU(N_c)\times SU(N_f)$.  Once the amplitude is larger than $\Lambda$, this
direction corresponds to
\begin{equation}
  Q = \left( 
    \begin{array}{cccc|c}
      0 & & & & 0\\
      & b^{1/N_c} & & & 0\\
      & & \ddots & & 0\\
      & & & b^{1/N_c} & 0
    \end{array} \right), \qquad \Qbar = 0.
\end{equation}
Again the potential goes up as $V_{IR} = N_c (\tm^2 - \mu^2)
|b|^{2/N_c}$ at large amplitudes.  Thus the vacuum of the theory will
settle at an intermediate scale where we lose calculability.

Since we did not truly minimize the potential, the two symmetry
breaking patterns above should be taken as candidates. One may hope
that once the theory begins to roll down one of these directions it
will `try to maintain as much of its symmetry' and stay along that
direction.  However, it is possible that the true minimum lies
elsewhere.

In summary, when $N_f=N_c+1$ and soft masses are added, the origin of
moduli space is unstable and thus some of the SQCD global symmetries
must be broken. We found two candidate directions. In the first case
the theory is not unstable right at the origin but becomes so near the
origin as long as $N_c$ is not too big. In this direction the chiral
symmetries are broken to the vector subgroup, and the meson VEVs as in
Equation~(\ref{eqn:flatdir}) show a color-flavor locked pattern for
$N_c$ of the flavors.  However, in the second direction, which is
unstable for any number of colors, some of the chiral symmetries
remain unbroken.  It appears that there is a first order phase
transition between the two directions as $\mu$ is increased.

In both cases baryon number is broken dynamically, consistent with the
result in non-supersymmetric QCD. Both of the candidate symmetry
breaking patterns differ from the previous results for
non-supersymmetric QCD as shown in Table~\ref{tab:qcd}.

\section{$N_c+2\le N_f < 3N_c$ -- Dual Region} \label{duality}

In this section we would like to analyze SQCD when $N_c+2\le N_f <
3N_c$.  Seiberg has shown~\cite{Seiberg:1994pq} that the original
$SU(N_c)$ ``electric'' theory is dual to a different ``magnetic''
gauge theory.  The dual theory consists of an $SU(\tnc)$ gauge theory,
where $\tnc = N_f - N_c$, with $N_f$ flavors of magnetic quarks $q_i$
and $\qbar_i$ along with a gauge singlet $M_{ij}$ and superpotential
$W=\frac1k q_i M_{ij} \qbar_j$.\footnote{Here we use $k$ instead of
the conventional $\mu$ in order to distinguish it from the chemical
potential.}  The two theories are obviously different in the UV, but
they are dual in the sense that they describe the same IR physics.
 
When $N_c+2 \le N_f < \frac{3}{2} N_c$, the so-called free magnetic
phase, the dual description is particularly useful because it is IR
free. In that case, the theory of dual quarks and the meson is
calculable in the IR and our goal is clear: we want to determine
whether the mass squared of the dual quarks is negative when a
chemical potential is added.

When $\frac32 N_c<N_f<3N_c$ both the electric and the magnetic
theories are asymptotically free but flow to an IR fixed point.  Our
analysis hinges on the ability to identify the correct low energy
degrees of freedom, their masses, and their charge under $U(1)_B$.
Once these are specified the effect of the chemical potential is
determined by gauge invariance and we can determine the stability of
the origin.  However, when the IR physics is strongly coupled the
quarks (electric or magnetic) may no longer be the correct degrees of
freedom. Therefore the fixed point value of the gauge coupling is
crucial in determining the reliability of our result.  As $N_f$ is
increased within this range the fixed point coupling for the electric
theory decreases whereas the fixed point coupling of the dual magnetic
theory increases. Once $N_f>3N_c$ the electric theory becomes IR-free
and the magnetic theory is asymptotically free.  Therefore, when
$\frac32 N_c<N_f<2N_c$ the magnetic theory is more reliable, whereas
for $2N_c<N_f<3N_c$ it is the electric theory we should be using. Near
$N_f=2N_c$ both theories are strongly coupled, so our analysis breaks
down in that region. Unfortunately we cannot determine the fixed point
couplings beyond the qualitative discussion above. Since all we can do
is calculate the mass squared of either the electric or magnetic
quarks we will therefore trust the result only in the two edges of the
conformal window where one of the descriptions is known to be weakly
coupled.

Before we discuss our results we will further comment on calculating
the critical chemical potential in the conformal window. Below we will
use exact supersymmetric results to calculate the masses of the IR
degrees of freedom, either electric or magnetic.  The mass calculated
in this manner is the effective mass at the scale $\Lambda$, where the
theory becomes conformal. If one would like to calculate the masses at
a lower scale $\omega$, the aproach to the fixed point must be
considered. The wavefunction renormalization of the quarks is
given by
\begin{equation}
Z=\left(\frac{\omega}{\Lambda}\right)^{\gamma_*}
\end{equation}
where $\gamma_*=b_0/N_f$ (or $\tilde{b}_0/N_f$ in the magnetic case)
is determined by the anomaly free $R$-charge. This decrease in $Z$
leads to an \emph{enhancement} in the effective IR masses
$m_{phys}=m/Z$. One may worry that the increase in the stabilizing
mass might make the breaking of $U(1)_B$ by a chemical potential more
difficult. However, this is not the case. We may write the effective
potential for the squrk at the scale $\omega\sim q$ as
\begin{equation}
V_{eff}(q)=m^2(q)|q|^2.
\end{equation}
If the effective mass $m^2(q)$ is negative for some scale (not
necesarily the IR) the potential becomes negative there and a
non-trivial minimum exists. Since the supersymmetric mass is only
enhanced below $\Lambda$, our best chance of achieving an instability
is for a scale of order~$\Lambda$.

The summary of our result for this case is as follows. When we add
supersymmetric masses and focus on the free magnetic phase we find
that a large enough chemical potential will make the dual quark masses
negative and break $U(1)_B$, provided a constraint on $k/\Lambda$ is
satisfied. Notice that this phase does not exist for $N_c=3$ so a
comparison with QCD results cannot be made.

For $N_f=\frac32 N_c$ there is no critical chemical potential within
our approximations.  For $\frac32 N_c < N_f < 2N_c$ a critical
chemical potential may exist if a certain restriction on $k/\Lambda$
is satisfied. However the naive expectation of $k\sim\Lambda$ does not
satisfy the restriction.  This is to be compared with the QCD result
that $U(1)_B$ is broken for $N_f=5$ which is at the bottom of the
conformal window ($N_f=1.66N_c$).

If we extrapolate our results to $N_f=2N_c$ we find that the critical
chemical potential needed to destabilize the magnetic quarks
approaches the electric quark mass. This shows that the theory
maintains its self duality for $N_f=2N_c$ when a chemical potential is
added, and might indicate that $U(1)_B$ can no longer be broken
dynamically in either description. However, this is the region where
we have the least control since both descriptions of the theory are
strongly coupled.

Extrapolating further to $2N_c<N_f<3N_c$ we again find that the mass
squared for the magnetic squarks may become negative with a sufficient
chemical potential (without requirements on $k$), however in this
region the electric quarks are more reliable. By our construction the
electric quarks do not condense dynamically and therefore $U(1)_B$ is
not broken for this case either.

When soft-masses are added in the UV theory we have control over the IR
masses only in the free magnetic phase. There baryon-number is broken
even without a chemical potential. Even though a finite chemical
potential leads to a ``stronger'' breaking of baryon number, it does
not lead to a phase transition like that found in other cases.  

In the conformal window, in addition to the difficulties described
above, the IR soft masses are uncalculable. Qualitatively we expect
that near $N_f=\frac32 N_c$ baryon number is dynamically broken by a
chemical potential and near $N_f=3N_c$ it is not. This will be argued
in Section~\ref{softmass-conformal}.

\subsection{Supersymmetric Mass} \label{susymass-duality}

First we will consider adding a supersymmetric mass for all the quarks
in the original theory in the form of a superpotential
$W_\mathrm{mass} = m_{ij}Q_i \Qbar_j$, with $m_{ij}=m\delta_{ij}$ as
before, which preserves the vector flavor symmetry. In the dual
theory, we thus have a superpotential
\beq
\label{eqn:fterm}
W_\mathrm{dual} = \frac1k q_i M_{ij}\qbar_j + \Tr (mM).
\eeq
The equations of motion for $M$ yield $q_i\qbar_j = m_{ji}$, which
does not admit any solutions because $q_i\qbar_j$ is at most a rank
$\tnc$ matrix, whereas $m_{ij}$ is chosen to have rank $N_f >
\tnc$. However, since we know that $\vevm \neq 0$ in the
electric theory, it suggests that the same must be true in the
magnetic theory. Then $\vevm$ becomes a mass term for the dual quarks
$q$, so we can integrate them out, leaving a pure $SU(\tnc)$
Yang-Mills theory with an additional gauge singlet $M$. The Yang-Mills
theory undergoes gaugino condensation, which gives a superpotential
term $W_{\lambda\lambda} = \tnc\tilde{\Lambda}_{IR}^3$. The usual
matching conditions give
$\tilde{\Lambda}_{IR}^{\tilde{b}_0'}=\tilde{\Lambda}^{\tilde{b}_0}
\det\langle M\rangle$, so together with the superpotential for $M$ we
find an effective superpotential
\beq
W_\mathrm{eff} = \tnc \left(\tilde{\Lambda}^{3\tnc-N_f}
\det \langle M\rangle  \right)^{1/\tnc} + \Tr (mM).
\eeq
Minimizing with respect to $M_{ij}$ gives an expression for $m_{ij}$
which, together with the duality relation $\Lambda^{b_0}
\tilde{\Lambda}^{\tilde{b}_0} = (-1)^{\tnc} k^{N_f}$, can be inverted to
give $\langle M_{ij}\rangle = (m^{-1})_{ij} \left(\Lambda^{b_0} \det m
\right)^{1/N_c}$, which is consistent with the assumption that
$\vevm\neq 0$ and matches the result in the electric theory.

\subsubsection{Free Magnetic Phase}

In the free magnetic phase we can estimate the size of $k$ as follows.
In the duality relation, $\Lambda$ is the scale where we expect the
electric theory to become strongly coupled, whereas $\tilde{\Lambda}$
is the high scale where the IR-free magnetic theory is strong. Thus it
is reasonable to identify these two scales, so that the theory is
described by the UV-free electric theory that flows down and becomes
strong around $\Lambda=\tilde{\Lambda}$ where the IR-free magnetic
theory takes over at strong coupling and subsequently flows to weak
coupling in the IR. With this identification, $k$ is also naturally
the same size as $\Lambda$. This argument also holds for the lower
edge of the conformal window where the magnetic theory is weak but for
a larger number of flavors the size of $k$ is unknown.

When we add a chemical potential the dual quarks will get a negative
mass-squared proportional to their baryon number, namely $N_c/\tnc$,
in addition to the mass from $\vevm$. So altogether,\footnote{This
  formula assumes that the K\"ahler potential $\int d^4 \theta Z_q q^*
  q$ for the dual quarks is canonical ($Z_q=1$).  If not, the
  normalization $Z_q$ appears in the formula by replacing $k$ by $Z_q
  k$.  In fact, it is known that scales $k$, $\Lambda$,
  $\tilde{\Lambda}$ are not physical, but rather depend on the
  normalization in K\"ahler potential \cite{holomorphy}.  This formula
  as well as the intuition $k \simeq \Lambda \simeq \tilde{\Lambda}$
  make sense in the {\it convention}\/ that the K\"ahler potential is
  canonical both for the electric quark in the UV limit and the
  magnetic quark in the IR limit.}
\beq
m^2_{q,\qbar} = \frac{1}{k^2}\vevm^2 - \left(\frac{N_c}{\tnc}\right)^2
\mu^2 = \frac{1}{k^2}\Lambda^{2b/N_c} m^{2(N_f/N_c - 1)} -
\left(\frac{N_c}{\tnc}\right)^2 \mu^2,
\label{eq:mq2}
\eeq
which leads to a critical chemical potential
\beq
\mu_c^2 = \left(\frac{\tnc}{N_c} \right)^2 \left(\frac{\Lambda}{k}
\right)^2 \left(\frac{m}{\Lambda} \right)^{2(N_f/N_c -1)} \Lambda^2.
\eeq
Thus for $\mu>\mu_c$ the dual squarks will have tachyonic masses,
signaling an instability. Because all the squarks get the same mass,
we expect them to all get the same VEV $x$ which, in order to satisfy
the $SU(\tnc)$ D-term equations, is of the following form in terms of
$\tnc\times N_f$ matrices:
\beq
q= \left( \begin{array}{cccccc} 
x &     &     &     & &\\
    & x &     &     & &\\
    &     & x &     & &\\
    &     &     &\ddots & & \end{array}\right)
\qquad \mbox{and} \qquad
\qbar= \left( \begin{array}{cccccc} 
\bar{x} &           &           &      & &\\
          & \bar{x} &           &      & &\\
          &           & \bar{x} &      & &\\
          &           &           &\ddots& &
\end{array}\right).
\eeq
Such a VEV completely breaks the $SU(\tnc)$ gauge group, which means
there is no residual $U(1)$ that could be combined with $U(1)_B$ to
give an unbroken $U(1)$, hence we conclude that $\mu$ exceeding the
critical chemical potential will break baryon number. But we still
need to check that such a value of $\mu$ is permitted within the
validity of our approximations.

We require $\mu_c < \Lambda$ so that the strong-coupling results can
be applied, which leads to an upper bound on $m/\Lambda$,
\beq
\frac{m}{\Lambda} <
\left(\frac{N_c}{\tnc}\frac{k}{\Lambda}\right)^{N_c/(N_f-N_c)}.
\label{eq:validity}
\eeq
The requirement that $\mu_c < m$ for stability in
the UV leads to another constraint:
\bea
\frac{m}{\Lambda} &>&
\left(\frac{N_c}{\tnc}\frac{k}{\Lambda}\right)^{N_c/(2N_c-N_f)}.
\eea
The magnetic theory is free in the infrared, so we are justified in
using the description in terms of dual quarks. Furthermore, the upper
bound on $m/\Lambda$ has a larger exponent than the lower bound this
time, which means we need $k/\Lambda > N_f/N_c-1$, which is a quite
reasonable constraint since the right hand side is always less than
$1/2$. As argued above, we expect $k$ and $\Lambda$ to be comparable.
This means that it is reasonable to expect $k/\Lambda > N_f/N_c-1$,
yielding a viable window where $\mu>\mu_c$ exists and all our
approximations remain valid.  Therefore there is a critical chemical
potential, of the form shown in the right diagram of
Figure~\ref{fig:mum-plot}, above which baryon number is broken.

\subsubsection{The Conformal Window}

In this case we obtain the same formula for the dual quark mass as in
Eq.~(\ref{eq:mq2}).  The validity requirement of $\mu_c < \Lambda$
also leads to the same constraint Eq.~(\ref{eq:validity}).  
On the other hand, the requirement that $\mu_c < m$ for stability in
the UV leads to three separate cases depending on the size of $N_f$:
\bea
\frac{m}{\Lambda} &>&
\left(\frac{N_c}{\tnc}\frac{k}{\Lambda}\right)^{N_c/(2N_c-N_f)}\ \ \
\mathrm{for}\ \ \ N_f<2N_c \\  
\frac{m}{\Lambda} &<&
\left(\frac{N_c}{\tnc}\frac{k}{\Lambda}\right)^{N_c/(N_f-2N_c)}\ \ \
\mathrm{for}\ \ \ N_f>2N_c \\
\mu_c &=& \frac{\Lambda}{k}m\ \ \ \mathrm{for}\ \ \ N_f=2N_c
\eea
We see that there are three situations. For $N_f=2N_c$, $\mu_c$ is
directly proportional to $m$ so we need $\Lambda/k < 1$. For more
flavors, $2N_c < N_f < 3N_c$, there are two different upper bounds, so
it is easy to choose a small $m$ that satisfies both. However, this
region is the upper part of the conformal window where the magnetic
theory is more strongly coupled, so our analysis in terms of the
magnetic quarks is unlikely to be valid. Instead it is more
appropriate to consider the electric theory where $U(1)_B$ is not
dynamically broken by construction, since we require $\mu<m$.

The more interesting situation is when $N_f < 2N_c$. In this case
$m/\Lambda$ is bounded both from above and from below. For
$N_f=\frac32 N_c$ the two bounds coincide, yielding $\mu_c=\Lambda$,
which doesn't leave any room for $\mu>\mu_c$ within our
approximations. For $\frac32 N_c < N_f < 2N_c$ the exponent of the
lower bound is larger than the exponent of the upper bound, so for
there to be any space between them, we need $(N_c/\tnc)(k/\Lambda) <
1$, which means $k/\Lambda < N_f/N_c - 1$, which is always less than
$1$ for this range of flavors. While not impossible to satisfy, the
naive expectation for $N_f$ close to $\frac32 N_c$ is that
$k/\Lambda\sim 1$.

We conclude that $U(1)_B$ remains unbroken in the conformal window.


\subsection{SUSY-Breaking Mass}

We now wish to add soft masses for the electric squarks in the UV.
Again we discuss the free magnetic phase and the conformal window
separately. 

\subsubsection{Free Magnetic Phase}
\label{susybreak-freemag}

Since the soft mass modifies the electric theory we must modify the magnetic
theory as well in order for the two theories to describe the same
moduli space.  In~\cite{Arkani-Hamed:1998wc} it was shown by arguments
similar to those of Appendix \ref{exactresults} that the duality is
maintained if soft masses are added to the magnetic quarks and the
meson. In the deep IR these soft masses take the form
\beq
\label{softmass-freemag}
m^2_M = 2\frac{3N_c-2N_f}{3N_c-N_f} \tm^2 \ \ \ \ \ \mathrm{and} \ \ \
\ \ m^2_{q,\qbar} = - \frac{3N_c-2N_f}{3N_c-N_f} \tm^2 -
\left(\frac{N_c}{\tnc} \right)^2 \mu^2.
\eeq
where we have included the contribution due to the chemical
potential. Note that the meson soft masses are positive, but as in the
previous case, the dual quarks $q$ and $\qbar$ get tachyonic masses.
Again, since the magnetic theory is free in the IR, its quarks and
meson are the appropriate degrees of freedom to analyze near the
origin of moduli space.

To analyze the symmetry breaking pattern first note that since the
$F$-term potential derived from Equation~(\ref{eqn:fterm}) does not
relate the expectation value of the dual quarks to the meson (or its
cofactor), the meson VEV will always vanish.  Representing the squark
and anti-squark as two $\tnc\times N_f$ matrices, the D-term equations
can be satisfied with a VEV of the form
\beq
\label{Dflat}
q= \left( \begin{array}{cccccc} 
x_1 &     &     &     & &\\
    & x_2 &     &     & &\\
    &     & x_3 &     & &\\
    &     &     &\ddots & & \end{array}\right)
\qquad \mbox{and} \qquad
\qbar= \left( \begin{array}{cccccc} 
\bar{x}_1 &           &           &      & &\\
          & \bar{x}_2 &           &      & &\\
          &           & \bar{x}_3 &      & &\\
          &           &           &\ddots& &
\end{array}\right) 
\eeq
with $|x_i|^2-|\bar{x}_i|^2=r$ where $r$ is a common constant for all $i$.

In order to find the vacuum we must minimize the rest of the potential
(the non D-terms) with respect to $x_i$ and the constant $r$. We can
assume without loss of generality that $r>0$.  The potential including
the soft terms is
\beq
\label{potentialx}
V=\sum_{i} |c_2|^2 |x_i|^2 \left( |x_i|^2+r \right)
+\sum_i m_q^2 \left(2|x_i|^2+r \right)
\eeq
where $m_q^2$ is negative as in
Equation~(\ref{softmass-freemag}). This is minimized for $x_i=0$ with
$r$ going to infinity due to the negative mass squared. Since we have
stabilized the theory for VEVs far away from the origin we expect $r$
to stabilize at some finite value of order $\Lambda$. The instability
of the origin exists even in the absence of a chemical potential since
the dual squark masses in Equation~(\ref{softmass-freemag}) are
negative. The case of $N_f=\frac32 N_c$ is an exception since then the
masses of Equation~(\ref{softmass-freemag}) vanish when $\mu$ is set
to zero, and $r$ is a modulus.  In that case an arbitrarily small
chemical potential will destabilize the origin and break $U(1)_B$.

Again we can only exclude the origin as a possible vacuum and point in
the direction the theory will roll away from the origin.  If the
vacuum indeed lies along this direction the symmetry breaking pattern
resembles the second flat direction for $N_f=N_c+1$ with soft
SUSY-breaking masses. Referring back to the quark VEVs in
Equation~(\ref{Dflat}) we see that this vacuum preserves the symmetry
$SU(\tnc)_{R+c}\times SU(N_f-\tnc)_R\times
SU(N_f)_L=SU(\tnc)_{R+c}\times SU(N_c)_R\times SU(N_f)_L$
\footnote{ Note that the $SU(N_c)$ symmetry here is a
\emph{flavor} symmetry.}. Negative $r$ would correspond to the
situation with left and right interchanged.  We see here that the free
magnetic phase exhibits some color-flavor locking in the sense that
the (dual) color gauge group and a subgroup of the flavor symmetry are
broken to the diagonal.

\subsubsection{The Conformal Window} \label{softmass-conformal}

In~\cite{Luty:1999qc} it was shown that, unlike the supersymmetric
masses that get enhanced, the soft masses in both the electric and the
dual magnetic theories flow to zero in the IR. The rate at which this
occurs is unknown and therefore we cannot determine whether the theory
reaches the fixed point.  Let us illustrate this qualitatively by
focusing on the electric description.  When the theory is weakly
coupled (near the Banks-Zaks fixed point~\cite{Banks:nn}) the soft
mass flows toward
zero very slowly so we can approximate it to be constant at its bare
value $\tm_0$. The conformal symmetry will be broken at the scale of
the soft mass. When the theory is more strongly coupled the soft mass
$\tm(\omega)$ is a more rapidly decreasing function of the scale
$\omega$. In this case the conformal symmetry will be broken at a
scale that is equal to the soft mass at that scale
$\omega\sim\tm(\omega)$. In this case the IR theory will not be
conformal nor supersymmetric, though the SUSY- breaking operator has
been suppressed.

Finally we may imagine that at yet stronger coupling the soft mass
$\tm(\omega)$ flows down with scale faster than the scale $\omega$
itself. In this case the condition $\omega\sim\tm(\omega)$ is never
satisfied and the theory reaches a truly conformal fixed point. Note
that even though supersymmetry was broken in the UV, it is restored in
the IR.  (See~\cite{Goh:2003yr} for an application of such a scenario
as a solution for the hierarchy problem.) However, as was pointed out
in~\cite{Luty:1999qc}, since we cannot calculate the rate of approach
to the fixed point we cannot determine whether this occurs.

>From the qualitative picture just described we can get a rough idea
about the size of the critical chemical potential for various numbers
of flavors within the conformal window. Near $N_f=\frac32 N_c$ when
the electric theory is strongest the IR soft mass is suppressed the
most and a smaller chemical potential is needed to break
$U(1)_B$. Since the suppressed mass is obviously smaller than the UV
mass, our assumption $\mu<\tm_0$ is satisfied and the $U(1)$ breaking
is indeed dynamical. If the last scenario of the previous paragraph
occurs, baryon number is broken for an arbitrarily small $\mu$.

As we add flavors and move up in the conformal window the suppression
weakens and the critical chemical potential rises. The naive
expectation is that it will reach $\tm_0$ near $N_f=3N_c$. Note that
this matches the values of the critical chemical potential obtained
outside of the conformal window: $\mu_c=0$ for $N_f=\frac32 N_c$ in
Section~\ref{susybreak-freemag}, and $\mu_c=m$ for $N_f\ge 3N_c$ where
the theory is IR free, by the arguments of Section~\ref{relBE}. The
rising trend for the critical chemical potential is also consistent
with the results for supersymmetric stabilizing masses in
Section~\ref{susymass-duality}.

\section{Conclusions}

Using known exact results for supersymmetric $SU(N)$ gauge theories
and soft masses, we have studied the effect of a baryon chemical
potential $\mu<\Lambda$ on the global symmetries of the ground state,
in particular $U(1)_B$. Our results are summarized in
Table~\ref{tab:sqcd}.
\renewcommand{\arraystretch}{1.2}
\begin{table}[!hb]
\begin{center}
\begin{tabular}{|c|c|c|c|}
\hline
& \multicolumn{2}{c|}{Unbroken Global Symmetries} \\
\cline{2-3}
\raisebox{2.0ex}{$N_f$} & SUSY Mass & SUSY-Breaking Mass \\ \hline \hline
$N_f<N_c$ & $SU(N_f)_V\times U(1)_B$ & $SU(N_f)_V\times U(1)_B$ \\ \hline
$N_f=N_c$ & $SU(N_f)_V$ & $SU(N_f)_V$ \\ \hline
\raisebox{-1.5ex}{$N_f=N_c+1$} & \raisebox{-1.5ex}{$SU(N_f)_V$} &
$SU(N_f-1) \times SU(N_f)$ \\[-1ex]
& & or $SU(N_f-1)_V$ \\ \hline
\raisebox{-1.5ex}{$N_c+2 \le N_f < \frac{3}{2} N_c$} &
\raisebox{-1.5ex}{$SU(N_f)_V$} & $SU(\tnc)_{R+c} \times
SU(N_c)_R \times SU(N_f)_L$ \\[-1ex]
& & or $(L \leftrightarrow R)$ \\ \hline
$\frac32 N_c < N_f < 2N_c$ & $SU(N_f)_V\times U(1)_B$ & $U(1)_B$ broken \\ \hline
$2N_c < N_f < 3N_c$ & $SU(N_f)_V\times U(1)_B$ & $U(1)_B$ broken? \\
\hline
\end{tabular}
\caption{Unbroken subgroup of the original $SU(N_f)_L\times SU(N_f)_R
\times U(1)_B$ global symmetry for SQCD with the addition of either
supersymmetric mass terms or soft SUSY-breaking masses. See the text
for assumptions made in each case.}
\label{tab:sqcd}
\end{center}
\end{table}

We have found that for $N_f<N_c$ a modified
$U(1)_B$ remains unbroken at low energy, whereas for $N_c \le N_f <
\frac32 N_c$ there is a critical chemical potential above which baryon
number symmetry is broken due to a combination of strong gauge
dynamics and the chemical potential. Notice that there is reasonable
agreement between the SUSY-preserving and SUSY-breaking
masses. Comparing with Table~\ref{tab:qcd}, we find that
$U(1)_B$ suffers the same fate as in non-supersymmetric QCD ($N_c=3$)
where it is unbroken in the 2SC phase ($N_f=2$) but broken in the CFL
phase ($N_f=3$). For SQCD with $N_f>3$ we also find interesting phases
that include possible $U(1)_B$ breaking. For $N_f=4$ we again find
$U(1)_B$ breaking, in agreement with the non-supersymmetric case. Our
results lend support to the results obtained using the variational
approach, even though the latter are derived in a very different
limit, namely $\mu\gg\Lambda$, and using very different methods.

\section*{Acknowledgments}
We would like to thank Markus Luty, Daniel Boyanovsky, and Paulo
Bedaque for helpful discussions. RH would like to thank Yael Shadmi
for additional discussions. This work was supported in part by the
Director, Office of Science, Office of High Energy and Nuclear
Physics, Division of High Energy Physics of the U.S. Department of
Energy under Contract DE-AC03-76SF00098 and in part by the National
Science Foundation under grant PHY-0098840. The work of HM was also
supported by Institute for Advanced Study, funds for Natural Sciences.

\appendix

\section{Harmonic Oscillator Example}
\label{HOexample}

The two-dimensional isotropic harmonic oscillator is an explicit
example which can be used to verify the formalism for treating the
chemical potential. We begin with the Hamiltonian for the 2D harmonic
oscillator and add a chemical potential for the ``number'' operator
$-L_z/\hbar$:
\beq
H - \mu N = \frac{p^2_x}{2m} + \frac{p^2_y}{2m} + \frac12 m\omega^2
(x^2+y^2) - \mu \frac{1}{\hbar}(y p_x - x p_y)
\eeq
which corresponds to the Lagrangian
\beq
L = \frac12 m \left[ \dot{x}^2 + \dot{y}^2 -
(\omega^2-\hat{\mu}^2)(x^2+y^2)-2\hat{\mu} (x\dot{y}-y\dot{x}) \right]
\eeq
where $\hat{\mu} \equiv \mu/\hbar$. Writing
$z=\frac{1}{\sqrt{2}}(x+iy)$ puts this in the form
\beq
L = m\left[ \dot{z}^* \dot{z} -(\omega^2-\hat{\mu}^2) z^* z
+i\hat{\mu} (z^*\dot{z}-z\dot{z}^*) \right]
\eeq
which (apart from the factor of $m$) is the Lagrangian for a
$(0+1)$-dimensional charged scalar field with ``mass'' $\omega$ and
nonzero chemical potential.

Apart from the similarity between this system and the field theory
that we are interested in, the virtue of this example is that we can
compare three separate computations for the partition function. First,
one can perform the sum over states in the grand canonical ensemble
directly.  This is most easily done by writing the Hamiltonian in
terms of two sets of creation and annihilation operators,
$H=(\hbar\omega-\mu)c^\dag c + \frac12 + (\hbar\omega+\mu)d^\dag d +
\frac12$, where $c=\frac{1}{\sqrt{2}}(a_x+ia_y)$ and
$d=\frac{1}{\sqrt{2}}(a_x-ia_y)$. The partition function then becomes
simply the product of two independent harmonic oscillators with
frequencies offset by $\pm \mu/\hbar$:
\beq
\mathcal{Z} = \sum e^{-\beta(H-\mu N)} = \frac{1}{2\sinh
\frac{\beta(\hbar\omega-\mu)}{2}} \frac{1}{2\sinh
\frac{\beta(\hbar\omega+\mu)}{2}}. 
\eeq
The same result is obtained by computing the grand canonical partition
function by an ``imaginary time'' path integral:
\beq
\mathcal{Z} = \Tr\ e^{-\beta(H-\mu N)} = \int \mathcal{D}x \mathcal{D}y\
e^{-\frac{m}{2\hbar}\int_0^{\beta\hbar} d\tau \left[
\dot{x}^2+\dot{y}^2 + (\omega^2-\hat{\mu}^2)(x^2+y^2) + 2i\hat{\mu}
(x\dot{y}-y\dot{x})  \right] }
\eeq
where here $\dot{x}=\del_\tau x$.

We can also perform the real-time path integral computation of the
quantum mechanical propagator, namely
\bea
\langle z_f,\,t_f|z_i,\,t_i \rangle &=& \int \mathcal{D}z^*
\mathcal{D}z\ e^{\frac{i}{\hbar} \int_{t_i}^{t_f} m \left[\dot{z}^*
\dot{z} -(\omega^2-\hat\mu^2)z^* z + i\hat\mu(z\dot{z}^*-z^*\dot{z})
\right]} \\
&=& e^{iS_c/\hbar} \frac{m\omega}{\pi i\hbar
\sin(\omega(t_f-t_i))}
\eea
where 
\beq
S_c = \frac{m\omega}{\sin\omega T}\left[ i(z_f^*z_i-z_f z_i^*)\sin\mu
T - (z_f^* z_i+z_f z_i^*)\cos\mu T + (|z_f|^2+|z_i|^2)\cos \omega T
\right]
\eeq
is the action over the classical path, and $T=t_f-t_i$. Taking $T
= -i\hbar\beta$ and $z_f=z_i=z$ followed by integration over $z$ gives
the grand canonical partition function:
\beq
\mathcal{Z}=\int dz \langle z|e^{-\beta(H-\mu N)}|z\rangle =
\frac{1}{2\sinh
\frac{\beta(\hbar\omega-\mu)}{2}} \frac{1}{2\sinh
\frac{\beta(\hbar\omega+\mu)}{2}}
\eeq
This matches the direct computation, verifying the form of the
effective Lagrangian in Minkowski space. This exercise demonstrates
how the $\mu^2$ term in the Lagrangian is necessary to produce the
correct result from a path integral calculation.

\section{Exact Results for Soft Masses} \label{exactresults}

The following is a review of~\cite{Arkani-Hamed:1998wc} and
\cite{Luty:1999qc}. We begin again with the SQCD Lagrangian
renormalized at a UV scale $\mu_{UV}$,
\begin{equation}
\label{Lsqcd}
\mathcal{L}_{UV}=\int d^4 \theta \sum_r 
\mathcal{Z}_r(\mu_{UV})Q_r^\dag Q_r+
\int d^2 \theta S(\mu_{UV}) \left(\mathcal{W}_\alpha
\mathcal{W}^\alpha + \mbox{h.c.}\right), 
\end{equation}
where the index $r$ labels the quark representations.  SUSY breaking
effects can be incorporated by promoting the couplings $\mathcal{Z}$
and $S$ to real and chiral superfields respectively as
follows\footnote{If there is no superpotential we can choose to absorb
the $\theta^2$ and $\bar{\theta}^2$ terms of $\mathcal{Z}$ into the
gluino mass through the anomalous symmetry of
Equations~(\ref{symanomalous}) and (\ref{symanomalous2}).}
\begin{eqnarray}
\label{spurions}
\mathcal{Z}_r&=&Z_r\left[1- \theta^2\bar\theta^2 \tm^2 
\right]
\nonumber\\
S&=&\frac{1}{g^2}\left( 1+\theta^2 \frac{m_\lambda}{2} \right).
\end{eqnarray}
Equation~(\ref{spurions}) incorporates a universal soft mass $\tm$ as
well as a possible gluino mass $m_\lambda$.

The theory possesses a spurious $U(1)_A$ symmetry under which
\begin{equation}
\label{symanomalous}
Q_r \to Q_r e^A \qquad \mathcal{Z}_r \to  \mathcal{Z}_r 
e^{-(A+A^\dag )}
\end{equation}
and the gauge coupling transforms anomalously
\begin{equation}
\label{symanomalous2}
S \to S - \frac{N_f}{8 \pi^2}A.
\end{equation}
The main point of~\cite{Arkani-Hamed:1998wc} and~\cite{Luty:1999qc} is
that this symmetry is respected by the strong dynamics that lead to
confinement. Since $\mathcal{Z}$ transforms as a gauge field under
this $U(1)_A$, it's couplings to the low energy degrees of freedom are
determined by this symmetry. This in turn determines the IR soft mass,
which is the highest component of the superfield
$\mathcal{Z}$. Furthermore, any physical quantity, related to either
the UV or the IR degrees of freedom, should be invariant under this
symmetry and also RG invariant.  It is therefore worthwhile to
construct a $U(1)_A$ and RG invariant object from $S$ and
$\mathcal{Z}$ since it will have physical meaning in both
descriptions. The only such object is
\begin{equation} \label{eqn:Idef}
I=\Lambda_h^\dag \mathcal{Z}^{\frac{2N_f}{b_0}} \Lambda_h 
\end{equation}
where $\Lambda_h=\mu_{UV}e^{-8 \pi^2 S(\mu_{UV})/b_0}$ is the
holomorphic dynamical scale and $b_0=3N_c-N_f$ is the one loop
beta-function coefficient.

Indeed, in Reference~\cite{Arkani-Hamed:1998wc} it was shown that for
UV-free theories the various components of $I$ are related simply to
the physical dynamical scale and the bare soft terms
\begin{eqnarray}
\label{Icomponents}
&&\left[I \right]_{\theta=\bar\theta=0}=\Lambda^2 \nonumber\\
&&\left[\ln I \right]_{\theta^2}=
\lim_{\mu_{UV}\to \infty}\frac{16\pi^2}{b_0}\frac{m_\lambda}{g^2} 
\nonumber\\
&&\left[\ln I \right]_{\theta^2\bar\theta^2}=
-\frac{2N_f}{b_0}\lim_{\mu_{UV}\to \infty}\tm^2_r
\end{eqnarray}
To find the effects of soft masses in the low energy theory, all we
need to do is write down the most general K\"{a}hler potential allowed
by the symmetries, including the anomalous $U(1)_A$, and determine the
coupling of $\mathcal{Z}$ to the kinetic terms. Since we only need
squark soft masses to stabilize the theory in the UV we will not add a
gaugino mass.

For example, consider the case with $N_f=N_c+1$. The K\"{a}hler
potential can be expanded around the origin and is
constrained by the symmetries and RG invariance to be
\begin{equation}
K=c_M\frac{\mathcal{Z}^2}{I}\Tr M^\dag M + c_B
\frac{\mathcal{Z}^{N_c}}{I^{N_c-1}} B_i^\dag e^{N_c V_B} B_i +
c_{\Bbar} \frac{\mathcal{Z}^{N_c}}{I^{N_c-1}}
\Bbar_i ^\dag e^{-N_cV_B} \Bbar_i +\cdots
\end{equation}
where $V_B$ is the background $U(1)_B$ gauge field. Note that since
this is an expansion around the origin, the potential we will derive
will only be valid for VEVs close to the origin that are much smaller
than $\Lambda$.

The soft masses for the mesons and baryons at some IR scale $\mu_{IR}$
will in general depend on the SUSY breaking VEVs of the spurion fields
$\mathcal{Z}$, $I$ and $V_B$ as well as $\mathcal{O}(1)$ numerical
coefficients $c_M$, $c_B$ and $c_{\Bbar}$. However, in
\cite{Arkani-Hamed:1998wc} it was shown that due to the IR-freedom of
the theory the soft masses don't depend on the coefficients $c_{M}$,
$c_B$, or $c_{\Bbar}$ in the deep IR, $\mu_{IR} \to 0$.  Using the
components of $\mathcal{Z}$ in Equation~(\ref{spurions}), $I$ in
Equation~(\ref{Icomponents}) and $V_B$ in Equation~(\ref{background})
and performing the $d^4\theta$ integral gives the soft masses for the
canonically normalized mesons and baryons.
Altogether we have:
\beq
\label{softmasses2}
\tm^2_M = \frac{2N_c-4}{2N_c-1} \tm^2 
\ \ \ \ \mathrm{and} \ \ \ \
\tm^2_{B,\Bbar} = \frac{2-N_c}{2N_c-1} \tm^2 - N_c^2 \mu^2.
\eeq
Notice that for $N_c \geq 3$ the meson mass-squared is positive, but
the baryons are tachyonic.

\bibliographystyle{unsrt}

\end{document}